\documentclass[journal]{IEEEtran}

\usepackage{latexsym, epsfig, tabularx}

\begin{document}

\title{{\Large \bf On the Capacity of Mobile Ad Hoc Networks with Delay Constraints}\thanks{This paper was presented in part to the IEEE CAS Workshop on Wireless Communications and Networking, Pasadena, California, September 2002. This research was supported by the National 
Science Foundation under Grant CCR-02-05214 and by the New Jersey Center for
Pervasive Information Technology.}}
\author{Cristina Comaniciu\footnote{This work was completed while the author was with 
Princeton University}, \hspace{2.0cm}H. Vincent Poor\\ 
Stevens Institute of Technology, \hspace{0.7cm} Princeton University \\
e-mail: {\tt ccomanic@stevens.edu, \hspace{0.5cm} poor@princeton.edu}}

\date{}
\maketitle

\thispagestyle{empty}
\pagestyle{empty}

\begin{abstract}
Previous work on ad hoc network capacity has focused primarily
 on source-destination throughput requirements for different
models and transmission scenarios, with an emphasis on delay tolerant
applications. In such problems, network capacity enhancement is achieved as a 
tradeoff with transmission delay. In this paper, the capacity of ad hoc
networks supporting delay sensitive traffic is studied. First, a general
framework is proposed for characterizing the interactions between the 
physical and the network layer in an ad hoc network. Then, CDMA 
ad hoc networks, in which advanced signal processing techniques such as 
multiuser detection are relied upon to enhance the user capacity, are analyzed. The network capacity is characterized using a combination of geometric arguments and large scale analysis, 
for several network scenarios employing matched 
filters, decorrelators and minimum-mean-square-error receivers.
Insight into the network performance for finite systems is also provided by 
means of simulations. Both analysis and simulations show a 
significant network capacity gain for ad hoc networks employing multiuser 
detectors, compared with those using matched filter receivers, as well as very 
good performance even under tight delay and transmission power requirements.

\end{abstract}

\begin{keywords}
ad hoc network, delay, capacity, CDMA, multiuser detection.
\end{keywords}
\section{Introduction}

A mobile ad hoc network consists of a group of mobile nodes that
spontaneously form temporary networks without the aid of a fixed 
infrastructure or centralized management. The communication between any two 
nodes can be either direct or relayed through other nodes (if the direct 
transmission causes too much interference in the network or consumes too much
power). Research on 
ad hoc networks has traditionally been focused on routing and medium 
access control, and only recently has  there been an increased interest in 
characterizing the capacity of such networks. We mention here a few landmark 
papers that analyze 
network capacity in terms of achievable throughput under different system 
models and assumptions \cite{tse_adh, gkumar, golds}. 
In \cite{golds}, the authors focus on fixed, finite networks and derive
capacity regions under various predefined transmission protocols, under the 
assumption of omniscient nodes. Alternatively, \cite{gkumar} and 
\cite{tse_adh} discuss the asymptotic throughput performance for fixed and 
mobile networks, respectively.
In \cite{gkumar}, the authors study the capacity of a fixed ad hoc network in
which the nodes' locations are fixed but randomly distributed. They  
prove that, as the number of nodes ($N$) per unit area increases, the 
achievable throughput between any randomly selected source-destination pair is 
on the order of $O(1/\sqrt{N})$.
In contrast to this somewhat pessimistic result,  \cite{tse_adh} shows that
exploiting mobility can result in a form of multiuser diversity and can 
improve the system capacity. The authors of \cite{tse_adh} propose a two-hop 
transmission 
strategy in which the traffic is first randomly spread (first hop) across as 
many relay nodes as possible, and then it is delivered (second hop) as soon 
as any of the relaying nodes is close to the destination.
The disadvantage of this scheme is that it involves large delays and therefore 
it is not suitable for delay sensitive traffic.
A capacity increase with mobility has also been noticed in \cite{gdas}, 
in which the capacity is empirically determined for a different network model 
that exploits spatial diversity. 

In this paper, we study the capacity of large mobile ad hoc networks carrying 
delay sensitive traffic. Because of tight delay requirements, we cannot take
advantage of mobility as in \cite{tse_adh}. To improve the capacity we
rely on advanced signal processing techniques such as multiuser detection, 
which can be implemented adaptively and blindly (e.g. \cite{wangpoor}).
 
We analyze the network for a given stationary distribution of the mobile 
nodes' locations with constraints on the maximum number of hops between any 
arbitrary source-destination pair.
Using similar arguments as in \cite{gkumar} we show that limiting the 
maximum number of hops for any given transmission also improves the 
source-destination throughput by limiting the additional transmissions for the 
relayed traffic. On the other hand, reducing the number of hops has a negative
impact on the capacity by increasing the interference level. Thus, for delay
sensitive traffic, the network capacity is interference limited and 
multiuser receivers can significantly improve the performance. 

We first propose a general framework for characterizing the interactions 
between the physical and the network layer in an ad hoc network. This is 
based on defining a link probability, which relates to the performance of both physical and network layers. We then focus on CDMA ad hoc networks, and 
determine link probability expressions at the network layer and physical layer. 

To derive the network capacity, we characterize the asymptotic network 
performance based on geometric considerations, and obtain a constraint on 
the probability of maintaining a link such that the network's 
diameter is $D$, as the number of nodes increases without bound.
As these results are asymptotic in nature, we also validate them through 
simulations for finite values of $N$. The network diameter represents
the longest shortest path between any two nodes, and consequently, is
the maximum number of hops required for transmission between any given 
pair of nodes. The link probability $p$ also characterizes the physical layer and is 
defined to be the probability that the signal-to-interference ratio can be 
maintained above the desired target.
We compute $p$ for different scenarios (Code-division multiple-access (CDMA) 
with random spreading codes and matched filter, minimum-mean-square-error 
(MMSE)  and decorrelating receivers) using an asymptotic analysis 
(both the number of nodes and the spreading gain are driven to infinity while 
their ratio is kept fixed) \cite{tse_hanl}.

The paper is organized as follows. In Section 2 we introduce the system model,
Section 3 presents the general cross-layer framework for analyzing ad hoc networks,
while Section 4 presents the asymptotic capacity derivation and Section 5 
presents simulations for finite systems. Finally, some 
concluding remarks are presented in Section 6.

\section{System Model}

We consider an ad hoc network consisting of $N$ mobile nodes, having a 
uniform stationary distribution over a square area, of dimension 
$b \times b$. The multiaccess scheme is direct-sequence CDMA (DS-CDMA) and 
three types of receivers are considered: the
matched filter (MF), the decorrelator, and the linear minimum mean squared 
error receiver (MMSE). All nodes use independent, randomly generated and normalized
spreading sequences of length $L$. For simplicity, we assume that all nodes 
transmit with the same power, $P_t$, and we define the signal-to-noise ratio 
(SNR), as the ratio between the transmitted power and the noise power: 
$SNR=P_t/\sigma^2$. As in \cite{mostofa}, we consider a transmitter oriented protocol,
in which each transmitting node has its own signature sequence. Although this
implementation yields more complex receivers and longer acquisition times, it
has very good capturing probabilities, allowing multiple packet reception at
the same receiver node. To avoid collisions, multiple concurrent 
transmissions from the same node are not allowed; instead transmissions from 
one node to multiple destination nodes are time multiplexed.  
We assume that all nodes are active at a given time (to transmit their own 
packets or relayed traffic), although the analysis can be easily extended to 
the case in which only a fraction $\beta$ of nodes are active (in which case 
interference is reduced by a factor of $1/\beta$).

The path loss model is usually characterized by three zones: the near field 
zone, the free space path loss zone and the excess path loss zone.
The near field zone extends to a distance of
\begin{equation}
d_1=\frac{2D_{max}^2}{\lambda},
\label{eq:nearfd}
\end{equation}

\noindent where $D_{max}$ is the largest dimension of the antenna, and $\lambda$ is the wavelength of the carrier.  
The signal attenuation in this zone is the highest and it is usually not
modeled for typical applications. For an antenna dimension of $D_{max}= \lambda /2$, the near-field zone extends to $d_1= \lambda /2$. In this paper we 
approximate the path loss 
model, and we assume that no reception is possible within distance $d<d_m$,
where $d_m = \lambda > d_1$.

For distances $d \geq d_m$, and $d \leq d_2=\frac{4 h_t h_r}{\lambda}$ 
($h_t$ and $h_r$ 
are the heights of the transmitter and receiver antennas, respectively), the 
free space propagation model applies. For antenna heights greater or equal to 
$1$ meter (a resonable value for ad hoc networks), and 3 GHz frequency 
($\lambda = 0.1$ meters), we have $d_2 \geq 40$ meters. Thus, since we are 
considering low range transmissions in the ad hoc networks, in our analysis we 
consider a free space propagation path loss model for which the received 
power is given as:
\begin{equation}
P_r = P_t^*G_tG_r \frac{\lambda^2}{(4 \pi d)^2}=P_t \frac{\lambda^2}{d^2} = P_t h,
\label{eq:freesp}
\end{equation}

\noindent where $P_t$ represents the above-defined transmitted power, which incorporates
also the transmitting and receiving antenna gains and the constant 
$1/(4\pi)^2$, and $h=\frac{\lambda^2}{d^2}$ is the link gain.

Although not specifically addressed in this paper, the analysis can be 
extended to consider a general path loss exponent $s>2$, which may be useful 
in characterizing the performance for long range ad hoc networks.

The traffic can be directly transmitted between any two nodes, or it can
be relayed through intermediate nodes. It is assumed that the end-to-end delay 
can be measured in the number of hops required for a route to be completed.
The quality of service (QoS) requirements for the ad hoc network are the bit 
error rate (mapped into a signal-to-interference ratio requirement: SIR), the 
average source-destination throughput ($T_{S-D}$), and the transmission delay. 
Both the throughput and the delay are influenced by the maximum number
of hops allowed for a connection and consequently, by the network diameter $D$.
Using arguments similar to those in \cite{gkumar}, a simplified computation 
shows that, if the number of hops for a transmission is $D$, then each node 
generates $D l(N)$ traffic for other nodes, where $l(N)$ represents the 
traffic generation rate for a given node.   Thus, the total traffic in the
network must meet the stability condition $Dl(N)N \leq NW/L$, where
$W$ is the system bandwidth. This implies that the average 
source-destination throughput that can be supported by the network must meet
the condition 
\begin{equation}
T_{S-D} \leq \frac{W}{LD}.
\label{eq:max_th}
\end{equation}

We note that the throughput $T_{S-D}(N) = l(N) \leq\frac{W}{LD(N)}$ is actually
dependent on the number of nodes in the network $N$, which influences the 
achievable network diameter.
For notation simplicity, for the remainder of the paper we denote $D(N)=D$ and 
$T_{S-D}(N)=T_{S-D}$, while keeping in mind that both quantities are in fact
dependent on $N$. 

In \cite{gkumar}, it was argued that although (\ref{eq:max_th})
shows that the throughput decreases with an increase in the number of hops
required, this does not account for the fact that if the range
of a node increases, more collisions occur and the throughput decreases.
In our case, increasing the transmission range for the CDMA network is
achieved as a result of improved physical layer reception (increased 
multi-packet reception capability), and thus directly yields increased 
network throughput for a reduced achievable network diameter. 

In terms of SIR requirements, a connection can be established between two 
nodes if the SIR is greater than or equal to the target SIR $\gamma$. The 
obtained SIR for a particular link is random due to the randomness of the 
nodes' positions.
To compute the probability of a connection between any two nodes we rely
on results developed in \cite{lkdistr} concerning the distribution of 
distances between any two nodes, when the nodes' locations are uniformly 
distributed in a rectangular area. In \cite{lkdistr}, an exact distribution 
for the distances is obtained, with the cumulative distribution function (CDF) given as\\

$P(d \leq bx)= $\\
\begin{equation}
\mbox{\hspace{0.8cm}}= \left\{\begin{array}{l} 0 ; \\ x^2(1/2x^2-8/3x+\pi) ; \\ 4/3\sqrt{x^2-1}(2x^2+1)- \\ -(1/2x^4+2x^2-1/3)+ \\+2x^2\left[\sin^{-1}(1/x)-\cos^{-1}(1/x)\right]; \\ 1  ; \end{array} \begin{array}{l} x<0 \\  0 \leq x \leq 1 \\   \\ \\ 1 \leq x \leq \sqrt{2} \\ x \geq \sqrt{2}. \end{array} \right.
\label{eq:dist1}
\end{equation}

\noindent It is also shown in \cite{lkdistr} that this model is very close to
a model in which the nodes are distributed according to a Gaussian 
distribution  having standard deviation $\sigma_1 = b/k$, with $k=3.5$. 
The CDF of $d$ under this new  model is given by
\begin{equation}
P(d \leq k\sigma_1 x) = 1-\exp\left(-\frac{k^2}{4}x^2\right), \ x \geq 0.
\label{eq:dist2}
\end{equation}

\noindent Equivalently, (\ref{eq:dist2}) can be expressed as:
\begin{equation}
F_d (y) = 1-\exp\left(-\frac{k^2}{4 {b}^2}y^2\right), \ y \geq 0. 
\label{eq:dist2f}
\end{equation}

\begin{figure}[ht]
\centerline{
\epsfxsize=2.6 in\epsffile{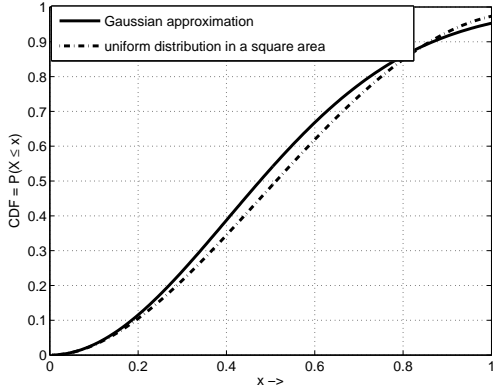}}
\caption{Gaussian approximation model: CDF}
\label{fig:distr}
\end{figure}

 \noindent The similarity between these two models is illustrated in Fig. 
\ref{fig:distr} for an example with $b=20$. For simplicity, we use the 
expression in (\ref{eq:dist2}) throughout the analysis, while the simulations 
rely on the actual uniform distribution over the square area.

We denote by $d_m = \lambda$ the minimum distance for reception and
by $d_M = \sqrt{2}b$ the maximum distance between two nodes 
(nodes uniformly distributed in a square area). We also define the normalized distances: $\delta_m=d_m /\lambda =1$, and $\delta_M = d_M/\lambda$. Hence, the 
link gain $h$ takes values in the interval $\left[1/\delta_M^2, \ 1/\delta_m^2\right]$ with high probability (e.g., according to the Gaussian model $P(d \leq \delta_m) \approx 8.5033 \times 10^{-4}$, and  $P(d \geq \delta_M)  \approx 0.0022$ for $\lambda = 0.1$ m and $b=6$).

As a consequence, the CDF for the link gain can be expressed as follows:
\begin{equation}
F_H(h) = 1-F_d\left(\lambda/\sqrt{h}\right)=\exp\left(-\frac{C}{h}\right), \ h > 0,  
\label{eq:cdf_lg}
\end{equation}

\noindent where $C= \frac{k^2}{4 {b}^2}\lambda^2$.

Taking the derivative of (\ref{eq:cdf_lg}) we obtain the probability
density function for the link gain:
\begin{equation}
f_H(h) = \frac{C}{h^2}\exp\left(-\frac{C}{h}\right), \ h \geq 0.  
\label{eq:pdf_lg}
\end{equation}

\noindent Using (\ref{eq:pdf_lg}) the mean link gain can be easily computed 
to be:
\begin{equation}
E_H \approx C\left[E_1(\delta_m^2C)- E_1(\delta_M^2C)\right],  
\label{eq:Eh}
\end{equation}

\noindent where $E_1(x)=\int_x^{\infty}\frac{1}{t}\exp(-t)dt$ is the 
exponential integral.

We define the network capacity to be the maximum number of nodes that can 
be supported such that both the SIR constraints and the delay constraints can
be met for any arbitrary source-destination pair of nodes. We map the delay
constraints into a maximum network diameter constraint $D$. 
In the following sections, we will characterize the ad hoc network 
asymptotic capacity for the case in which the number of nodes and the 
spreading gain go to infinity, while their ratio is fixed. 

\section{General Cross-Layer Framework for Ad Hoc Networks\label{sec:gen}}

We start our discussion by characterizing the interactions
between the physical and network layers for a wireless ad hoc network. 
The interactions between layers can be characterized based on a 
cross-coupling element, which represents the information shared between 
layers. This cross-coupling element is essentially related to the 
quality of the links. Since wireless systems are primarily interference 
limited, the physical layer transmission and reception parameters (e.g. 
transmitted powers, receiver design) influence the link quality, and
consequently the cross-coupling information. 
On the other hand, the links constitute the basic element to construct
routing graphs that are used to optimize routing at the network layer. 
Minimum cost routing may rely on cost definitions that include the 
shared links' quality information. 

\begin{figure}[ht]
\centerline{
\epsfxsize=1.4 in\epsffile{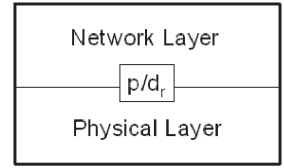}}
\caption{General cross-layer model for ad hoc networks}
\label{fig:genfr}
\end{figure}

A question that arises is: what is an appropriate link quality metric to
 serve as a cross-coupling element?
We note that there is no unique definition for the shared information measure
across layers. We suggest that two appropriate selections for the 
cross-coupling element are the reliable transmission range $d_r$, and 
the link availability probability $p$ (see Figure \ref{fig:genfr}). 

The reliable transmission range is defined as the maximum source-destination 
transmission distance for which the BER target can be met. We can easily 
see that improvements at the physical layer will yield an extended 
transmission range, and consequently higher $d_r$. This translates into 
longer links being available for routing at the network layer. If a network 
delay is imposed, which requires a minimum hop solution, a certain constraint 
on the desired value of $d_r$ can be enforced. 

\begin{figure}[ht]
\centerline{
\epsfxsize=2.6 in\epsffile{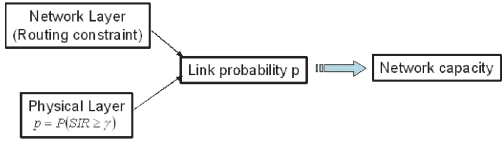}}
\caption{Link probability and network capacity}
\label{fig:p_ct}
\end{figure}

The other link quality metric proposed, the link probability constraint $p$,
is defined as the probability that a link is available for transmission, 
i.e., meets the SIR target constraints. We can see that this measure is more
generally applicable than the previous one, as it relates to link quality 
variations that are not necessarily distance based.
As a possible example, a power controlled network (equal SIR for all links)
may result in different achievable target SIRs for the links as a function
of the level of interference in the network. The network delay will 
include the effect of retransmissions for errored packets. For this 
case, at the network level, an average delay constraint as well as a variance
delay constraint may be imposed (to model delay and jitter), which coupled
with the physical layer characteristics will determine the network capacity
(see Figure \ref{fig:p_ct}). In a similar context, this measure may be more 
suitable for power controlled users in fading environments, where the link 
quality varies with the fading process, irrespective of distance.
Furthermore, the link availability model fits naturally with any random access
based system model with or without multipacket reception, where the link 
availability can be translated into the probability of success for the
current transmission on a particular link.
We also note that the two information sharing measures are related for
simple system models, and a relationship between them can be 
determined.

\begin{figure}[ht]
\centerline{
\epsfxsize=2.6 in\epsffile{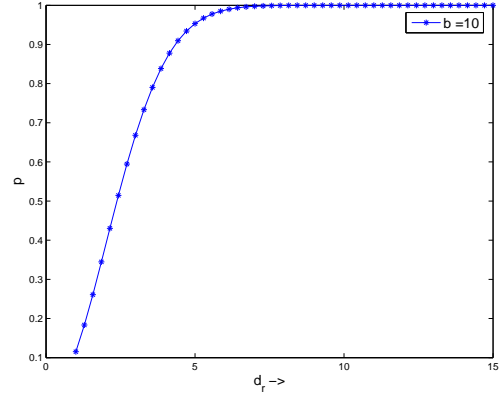}}
\caption{Cross-layer coupling metrics equivalence}
\label{fig:equiv}
\end{figure}

In this work, we consider only distance based fading, and we ignore the effect
of retransmissions on the packet transmission delay. Consequently, the
reliable transmission range metric $d_r$ is suitable to describe the 
cross-coupling between layers. However, since the link availability 
metric $p$ is suitable to model a larger array of scenarios, we will mostly
use $p$ in our derivation, with the understanding that, for our system model,
both metrics are equivalent.
In Figure \ref{fig:equiv} we show the equivalence between these two metrics, 
which can be expressed analytically as:

\begin{equation}
p = P\left(d\leq d_r\right)= F_d(d_r) = 1-\exp\left(-\frac{k^2}{4b^2}{d_r}^2 \right).
\end{equation}

\section{Asymptotic Capacity for CDMA Ad Hoc Networks }

\subsection{Physical Layer Performance\label{section:phy}}

Based on the framework presented in Section \ref{sec:gen}, we determine the 
number of active nodes that can be supported by the network, given a link
probability constraint $p$. The link probability $p$ is affected by the level 
of interference in the network and thus it will be very sensitive to the 
choice of the receiver. 

We start our capacity analysis by considering the
ideal synchronous transmission case, then we discuss  performance results
for the asynchronous case. The asymptotic capacity is derived 
for three types of receivers: the matched filter, the decorrelator and the 
linear MMSE detector.

\subsubsection {Synchronous Transmission}

\vspace{.2cm}
\indent {\bf Matched Filter}

The SIR condition for an arbitrary node $i$ using a matched filter receiver 
in a network with random, normalized spreading sequences can be expressed as:\\
\vspace{.2cm}

$SIR_{i}=\frac{P_th_i}{\sigma^2+\frac{1}{L}\sum_{j=1,\ j \neq i}^{N}P_th_j}$

\vspace{.2cm}
\begin{equation}
\mbox{\hspace{-1.3cm}}=\frac{h_i}{SNR^{-1}+\frac{1}{L}\sum_{j=1,\ j \neq i}^{N}h_j} \geq \gamma.
\label{eq:SIR_mf}
\end{equation}

Denoting by $\alpha$ the fixed ratio $N/L$ and letting the number of nodes and the spreading gain go to infinity, by using the law of large numbers \cite{prob_yg}, it follows that:
$\frac{1}{L}\sum_{j=1,\ j \neq i}^{N}h_j \rightarrow \alpha E_H$, with $E_H$
computed as in (\ref{eq:Eh}).

According to our framework, we derive the link probability $p$, such that physical
layer QoS requirements are met, i.e., the link meets its target SIR with probability $p$:
\begin{equation}
P\left(H \geq \gamma SNR^{-1}+ \alpha \gamma E_H \right) =  P\left(H \geq T_{MF} \right)=p.
\label{eq:defpmf}
\end{equation}

\noindent Using the notation $T_{MF}= \gamma SNR^{-1}+ \alpha \gamma E_H$,
the link probability condition renders an $SNR$ condition
\begin{equation}
\gamma SNR^{-1}+ \alpha \gamma E_H = T_{MF} \Rightarrow SNR = \frac{\gamma}{T_{MF}- \alpha \gamma E_H},
\label{eq:mfconstr2}
\end{equation}

\noindent where $T_{MF}$ can be derived using (\ref{eq:cdf_lg}) as follows
\begin{equation}
p= 1-F_H(T_{MF})=1-\exp\left(-C \frac{1}{T_{MF}}\right);
\label{eq:defpmf1}
\end{equation}

\noindent or
\begin{equation}
T_{MF}=\frac{C}{\log \left(\frac{1}{1-p}  \right)}.
\label{eq:mfconstr1}
\end{equation}

\noindent Equation (\ref{eq:mfconstr2}) implies that a positive power solution
exists if and only if 
\begin{equation}
\alpha_{MF} < \frac{T_{MF}}{\gamma E_H}= \frac{\frac{C}{\log \left(\frac{1}{1-p}\right)}}{\gamma E_H}.
\label{eq:capmf1}
\end{equation}

\noindent For ad hoc networks, it is most likely that the mobile nodes are
energy limited such that we assume that a maximum power transmission limit
$\bar{P_t}$ is imposed. Denoting $SNR_c=\bar{P_t}/\sigma^2$, the ad hoc network
capacity becomes:
\begin{equation}
\alpha_{MF} \leq \frac{T_{MF}}{\gamma E_H} - \frac{1}{E_H SNR_c}= \frac{\frac{C}{\log \left(\frac{1}{1-p}\right)}}{\gamma E_H} - \frac{1}{E_H SNR_c}.
\label{eq:capmf2}
\end{equation}

\vspace{.2cm}
\indent {\bf Decorrelator}

According to results presented in \cite{tse_hanl}, the SIR of an 
arbitrary node in an asymptotically large network using decorrelating 
receivers can be expressed as:
\begin{equation}
SIR_{d} = \left\{ \begin{array}{l} \frac{P_th(1-\alpha)}{\sigma^2}, \\ 0 \end{array} \begin{array}{r} \mbox{\hspace{2cm}} \alpha < 1, \\ \alpha \geq 1 \end{array} \right.
\label{eq:sirdec}
\end{equation}

\noindent Thus, if no power constraints are imposed, the network capacity region
is
\begin{equation}
\alpha_d < 1.
\label{eq:capdec1}
\end{equation}

\noindent If power constraints are imposed, and $SNR \leq SNR_c$ ($SNR_c$ is the  maximum SNR allowed), the physical layer constraint can be expressed as:
\begin{equation}
P\left(H \geq \frac{\gamma}{SNR(1-\alpha)} \right)=p. 
\label{eq:cond_dec1}
\end{equation}

\noindent If we define $T_d = \frac{\gamma}{SNR(1-\alpha)}$, the feasibility 
condition becomes 
\begin{equation}
SNR=\frac{\gamma}{T_{d}(1-\alpha)} \leq SNR_c. 
\label{eq:cond_dec2}
\end{equation}

\noindent Imposing a network constraint on the $T_d$ value, $T_d = \frac{C}
{\log \left(\frac{1}{1-p}\right)}$, the asymptotic capacity region for a network using 
decorrelating receivers and having transmission power constraints is given as
\begin{equation}
\alpha_d \leq 1- \frac{\gamma}{T_d SNR_c}=1- \frac{\gamma}{\frac{C}{\log \left(\frac{1}{1-p}  \right)} SNR_c}. 
\label{eq:cap_decc}
\end{equation}

\indent {\bf MMSE Detector}

To derive the asymptotic ad hoc network capacity we first express the SIR 
ratio for an arbitrary node $i$ in a large network using MMSE receivers, 
as in \cite{tse_hanl}:

\vspace{.2cm}
$SIR_i=\frac{P_th_i}{\sigma^2+\frac{1}{L}\sum_{j=1, j \neq i}^{N}\frac{P_th_iP_th_j}{P_th_i+P_th_j SIR_i}}=$

\begin{equation}
\mbox{\hspace*{-0.9cm}}=\frac{h_i}{SNR^{-1}+\frac{1}{L}\sum_{j=1, j \neq i}^{N}\frac{h_ih_j}{h_i+h_j SIR_i}}.
\label{eq:sirmmse}
\end{equation}

\noindent Imposing the QoS condition: $SIR_i \geq \gamma$, $\forall i=1,2,...,N$, (where $\gamma$ is the target SIR), we have   
\begin{equation}
SIR_i \geq \frac{h_i}{SNR^{-1}+\frac{1}{L}\sum_{j=1, j \neq i}^{N}\frac{h_ih_j}{h_i+h_j \gamma}} = \gamma.
\label{eq:sirmmse1}
\end{equation}

\noindent Denoting $\alpha = N/L$ and letting the spreading gain and the number of nodes go to infinity we can apply the law of large numbers, such that,\[\frac{1}{L}\sum_{j=1,\ j \neq i}^{N}\frac{h_i h_j}{h_i+h_j\gamma} = \alpha \frac{1}{N}\sum_{j=1,\ j \neq i}^{N}\frac{h_i h_j}{h_i+h_j\gamma}\rightarrow  \alpha E[H|h_i], \]

\noindent where we used the notation $E[H|h_i]$
to denote the normalized conditional average interference (normalized to the
number of nodes per dimension). It is shown in the Appendix that $E[H|h_i]$
can be expressed as: 

\vspace{0.3cm}

$E[H|h_i]=$
\begin{equation}
=C\exp\left(\frac{C \gamma}{h_i}\right) \left[E_1\left(\delta_m^2C+\frac{C \gamma}{h_i}\right)-E_1\left(\delta_M^2C+\frac{C \gamma}{h_i}\right)  \right]. 
\label{eq:Ehh}
\end{equation}

\noindent Thus, the link probability constraint becomes
\begin{equation}
P \left(H \geq \gamma SNR^{-1} + \alpha \gamma E[H|h] \right)=p.
\label{eq:lcmmse}
\end{equation}

\begin{figure}[ht]
\centerline{
\epsfxsize=2.6 in\epsffile{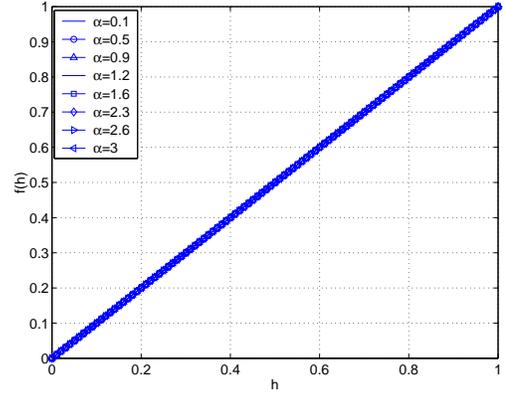}}
\caption{SIR condition monotonicity}
\label{fig:func}
\end{figure}

\noindent We define the function $f(h)= h - \gamma SNR^{-1} - \alpha \gamma E[H|h]$ and we plot it in Fig. \ref{fig:func}. We observe that $f(h)$ is a
monotonically increasing function of $h$ for the region of interest, and thus
we can express the condition (\ref{eq:lcmmse}) as 
\begin{equation}
P \left(H \geq T_{MMSE} \right)=p.
\label{eq:lcmmse1}
\end{equation}

Equation (\ref{eq:lcmmse1}) has the same solution as in the previously analyzed
cases, and the physical layer constraint becomes
\begin{equation}
SNR = \frac{\gamma}{T_{MMSE}-\alpha \gamma E[H|h = T_{MMSE}] }.
\label{eq:phmmse}
\end{equation}

\noindent A positive transmitting power solution exists if and only if
\begin{equation}
\alpha_{MMSE} < \frac{T_{MMSE}}{\gamma E[H|h = T_{MMSE}]};
\label{eq:capmmse1}
\end{equation}

\noindent or equivalently,
\begin{equation}
\alpha_{MMSE} < \frac{\frac{C}{\log \left(\frac{1}{1-p}  \right)}}{\gamma C \left( \frac{1}{1-p} \right)^{\gamma}\left[E_1\left( \zeta_m \right)-E_1\left(\zeta_M \right)  \right]},
\label{eq:capmmse2}
\end{equation}

\noindent where $\zeta_m = \delta_m^2C+ \gamma \log \left(\frac{1}{1-p}  \right)$, and $\zeta_M =  \delta_M^2C+ \gamma \log \left(\frac{1}{1-p}  \right)$.

\noindent If power constraints are imposed, the capacity region becomes
\begin{equation}
\alpha_{MMSE} \leq \frac{T_{MMSE}}{\gamma E[H|h = T_{MMSE}]}-\frac{1}{E[H|h=T_{MMSE}]SNR_c};
\label{eq:capmmse1c}
\end{equation}

\noindent or equivalently,
\begin{equation}
\alpha_{MMSE} \leq \frac{\frac{C}{\log \left(\frac{1}{1-p}  \right)}}{\gamma C \left( \frac{1}{1-p} \right)^{\gamma}\left[E_1\left(\zeta_m \right)-E_1\left(\zeta_M \right)  \right]} - 
\label{eq:capmmsef2}
\end{equation}
\[\mbox{\hspace{2cm}} - \frac{1}{ C \left( \frac{1}{1-p} \right)^{\gamma}\left[E_1\left(\zeta_m \right)-E_1\left(\zeta_M \right)  \right]SNR_c}.\]

\begin{figure}[ht]
\centerline{
\epsfxsize=2.6 in\epsffile{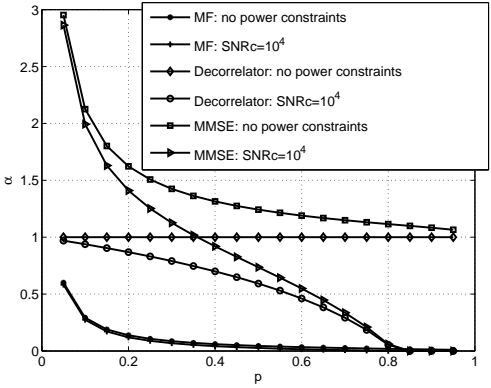}}
\caption{Physical layer capacity for given link probability constraint: synchronous transmission}
\label{fig:cap_ph}
\end{figure}

Figure \ref{fig:cap_ph} illustrates the physical layer capacity as a function
of the link probability constraint for the three receivers considered, and 
with or without power constraints. For the power-constrained case, a maximum 
transmission power of $\bar{P_t}= 10^4 \sigma^2$ is considered for this example. 
A target SIR $\gamma = 5$ is imposed.

From Figure \ref{fig:cap_ph} we can observe that there is a significant 
capacity advantage if multiuser receivers are used, and conversely, for given 
capacity requirements, substantial power savings can be achieved by networks 
using multiuser receivers. As expected, the MMSE receiver performs the best 
due to its property of maximizing the SIR.
For higher transmission rates and lower delay requirements (translated into a
higher link probability constraint) using the matched filter is not feasible.

\subsubsection{Asynchronous Transmission}

Although often not a practical scenario, the above synchronous analysis is 
very useful as a performance benchmark. Moreover, the analysis can usually be extended
to the asynchronous case by considering an equivalent synchronous system with 
more interferers. To extend our results to asynchronous ad hoc 
networks, we rely on SIR convergence properties developed in \cite{tse:asy}.
According to results presented in \cite{tse:asy}, the matched filter 
performance in the asynchrounous case has the same asymptotic performance as 
for the synchronous case. Also, if the observation window is infinite, the 
decorrelator and the MMSE performance does not change either. However, for the
``one-shot'' detection approach, the achieved SIR for both the decorrelator and
the MMSE detector degrades for the asynchronous case. Although exact capacity values are difficult to derive, in \cite{tse:asy} the authors present very tight lower bounds on the achievable SIR for both 
the decorrelator and the MMSE detector in the asynchronous case, under the 
simplifying assumption that the nodes are chip-synchronous. They also showed 
by means of simulations that the chip-synchronous scenario provides conservative 
estimates for a truly asynchronous system.

\vspace{.1cm}
\indent {\bf Decorrelator}

According to \cite{tse:asy}, the SIR for the decorrelator can be approximated
as  

\begin{equation}
SIR_{d} = \left\{ \begin{array}{l} \frac{P_th(1-2 \alpha)}{\sigma^2}, \\ 0 \end{array} \begin{array}{r} \mbox{\hspace{2cm}} \alpha < 1/2, \\ \alpha \geq 1/2 \end{array} \right.
\label{eq:sirdec_asy}
\end{equation}

\noindent Therefore, the capacity results from the previous subsection can be
extended straightforwardly to 
\begin{equation}
\alpha_d < 1/2,
\label{eq:capdec1asy}
\end{equation}

\noindent when no power constraints are imposed.

If power constraints are imposed ($SNR \leq SNR_c$), we derive the capacity region as
\begin{equation}
\alpha_d \leq  \frac{1}{2}- \frac{\gamma}{\frac{2C}{\log \left(\frac{1}{1-p}  \right)} SNR_c}. 
\label{eq:cap_decc_asy}
\end{equation}

\vspace{.2cm}
\indent {\bf MMSE Detector}

To characterize the capacity of an asynchronous ad hoc network using MMSE 
receivers, we rely on the lower bound obtained for the achievable SIR in
an asymptotic system for symmetric delay distributions \cite{tse:asy}:
\begin{equation}
SIR_i = \frac {P_i}{\sigma^2 + \alpha E_P E_{\tau}\{I(\tau P, P_i, SIR_i)+I((1-\tau)P, P_i, SIR_i)\}},
\label{eq:sir_mmse_asy}
\end{equation}

\noindent where $\tau$ is a random variable that characterizes the delay associated with an arbitrary node. 
\noindent Since the received power $P$ can be expressed as $P=P_t h$, 
for equal transmit powers for all nodes,  (\ref{eq:sir_mmse_asy}) becomes

\vspace{0.2cm}
$SIR_i =$
\vspace{0.2cm}
\begin{equation}
=\frac {h_i}{SNR^{-1} + \alpha E_H E_{\tau}\{I_{\tau} +I_{(1-\tau)}\}},
\label{eq:sir_mmse_asyh}
\end{equation}

\noindent where we used the notations: $I_{\tau}=I(\tau P_th, P_th_i, SIR_i)$, and  $I_{(1-\tau)}=I((1-\tau)P_th, P_th_i, SIR_i)$.

It is straightforward to see (using a similar derivation as in the
appendix) that $\alpha E_H E_{\tau}\{I(\tau P_th, P_th_i, SIR_i)+I((1-\tau)P_th, P_th_i, SIR_i)\}$ can be expressed as 

\vspace{0.2cm}
$\mbox{\hspace{-.5cm}}\alpha E_{\tau}\left[E[H|h_i,\tau] \right] =\alpha E_{\tau}\left [ C \tau \exp\left(\frac{C \gamma \tau}{h_i}\right) \left[E_1\left(\xi_m^{\tau}\right)-E_1\left(\xi_M^{\tau}\right)\right] + \right. $\\
\vspace{0.2cm}
\begin{equation}
 \left. + C (1-\tau) \exp\left(\frac{C \gamma (1-\tau)}{h_i}\right) \left[E_1 \left( \xi_m^{(1-\tau)} \right) -E_1\left(\xi_M^{(1-\tau)}\right) \right] \right],
\label{eq:sir_mmse_asyh2}
\end{equation}

\noindent where we used the notations $\xi_m^{(\tau)}=\delta_m^2C+\frac{C \gamma \tau }{h_i}$, $\xi_M^{(\tau)}=\delta_M^2C+\frac{C \gamma \tau}{h_i}$, $\xi_m^{(1-\tau)} = \delta_m^2C+\frac{C \gamma (1-\tau) }{h_i}$, and $\xi_M^{(1-\tau)} = \delta_M^2C+\frac{C \gamma (1-\tau)}{h_i}$.
\vspace{0.2cm}

$E_{\tau}\left\{E[H|h_i,\tau] \right\}$ can be determined using numerical integration. For our example, we have considered $\tau$ to be a uniform random variable taking values in the interval $[0, \ 1]$. 

\begin{figure}[ht]
\centerline{
\epsfxsize=2.6 in\epsffile{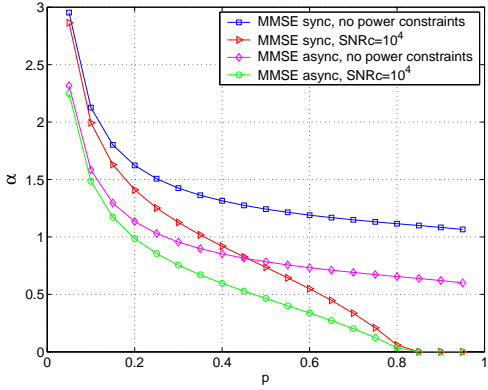}}
\caption{Capacity comparisons for ad hoc networks with MMSE receivers: synchronous versus asynchronous transmission}
\label{fig:mmse_asy}
\end{figure}

Using an identical derivation for the network capacity as for the synchronous 
case, all the capacity formulas hold with $E[H|h_i]$, replaced by $E_{\tau}\left\{E[H|h_i,\tau] \right\}$.
In Figure \ref{fig:mmse_asy}, we illustrate capacity comparisons between 
networks using MMSE receivers in the synchronous and the asynchronous cases.

\subsection{Network Layer Performance}

The overall network capacity is determined such that both physical layer and
network layer QoS requirements can be met. In the previous section we have
determined the maximum number of active nodes that can be supported 
simultaneously by the network, as a function of a link probability constraint $p$. 
In this section, we use geometric arguments to determine the dependence of 
the link probability on the network diameter constraint (which is a measure of the 
delay constraint).

\begin{figure}[ht]
\centerline{
\epsfxsize=2.6 in\epsffile{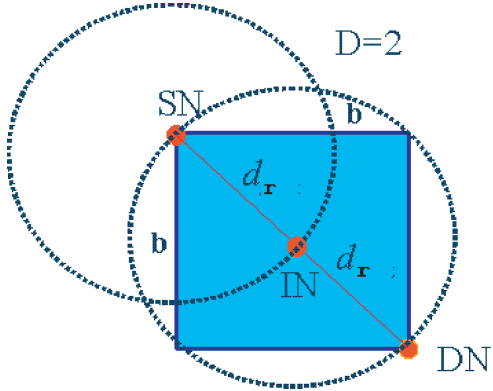}}
\caption{Network diameter constraint}
\label{fig:net_cap}
\end{figure}

We consider the asymptotic case, in which we have an infinite number of nodes
in the considered square area. The number of nodes is uniformly distributed,
and we ignore the edge effects: the square area can be considered to be a part
of a multiple cell layout. It can be seen in Figure \ref{fig:net_cap} that the
worst case distance is obtained when the source and destination nodes are on
the opposite vertices of the square.

Consider now a diameter restriction of $D=2$, as in Figure \ref{fig:net_cap}. In order to be able to transmit from the source node (SN) to the destination node (DN) using only one intermediate node (IN), the reliable transmission range
$d_r$ should be equal to the distance between the source node and intermediate node, and also between the intermediate node and the destination node, respectively: 
\begin{equation}
d_r = \frac{b\sqrt{2}}{2}.
\label{eq:D2}
\end{equation}
\noindent For a generic value of $D$, the network QoS constraint becomes
\begin{equation}
d\leq d_{r}=\frac{b\sqrt{2}}{D}.
\label{eq:Dgen}
\end{equation}

Thus, a link may be used for routing with probability
\begin{equation}
p=F_d(d_{r})=1-\exp\left(-\frac{C}{\lambda^2}\frac{2b^2}{D^2}\right).
\label{eq:pnetf}
\end{equation}

\begin{figure}[ht]
\centerline{
\epsfxsize=2.6 in\epsffile{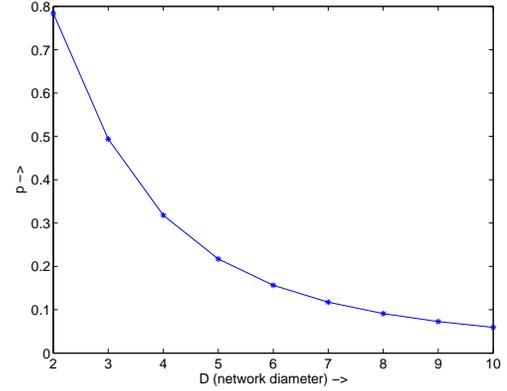}}
\caption{Link probability requirement}
\label{fig:p_net}
\end{figure}

\noindent Figure \ref{fig:p_net} illustrates the link probability values 
required for various network diameter constraints. The case $D=1$ is trivial, 
as $p \approx 1$ (the approximation is due to approximations in the 
derivation of distributions for link distances).

We note that the assumption that we will always find a relaying node at the required distance is optimistic for finite networks, whereas the assumption the the link length is equal 
to the diagonal is a pessimistic assumption. Simulaton results will be presented to illustrate 
how accurate this analysis is for finite networks.

\subsection{User Capacity}

In the previous sections, we derived link probability ($p$) expressions to 
reflect both physical layer and network layer QoS constraints. Based on these 
expressions, we can now derive the user capacity for the network 
(see Figure \ref{fig:p_ct}).

More specifically, (\ref{eq:pnetf}) gives the network probability constraint $p$ to be 
substituted into all of the user capacity equations derived in Section \ref{section:phy}.

\noindent Furthermore, we can see that the network QoS condition imposes that a reliable transmission is possible within a radius $d_{r}$ of the transmitting node:
\begin{equation}
p=P(H \geq T) = P\left(d_{r} \leq \frac{\lambda}{\sqrt{T}}\right),
\label{eq:dmaxc}
\end{equation}

\noindent where the threshold $T$ depends on the particular receiver structure 
used.

Based on this observation, an alternate simple way to derive the capacity is 
to substitute $h = \lambda^2/d_{r}^2$ in the physical layer capacity conditions and solve
for the number of users. Alternatively, from (\ref{eq:dmaxc}) and (\ref{eq:Dgen}), 
we can obtain a threshold requirement of
\begin{equation}
T = \frac{\lambda^2 D^2}{2b^2},
\label{eq:thr}
\end{equation}

\noindent which can then be used to determine the network capacity.

\begin{figure}[ht]  
\centerline{
\epsfxsize=2.6in\epsffile{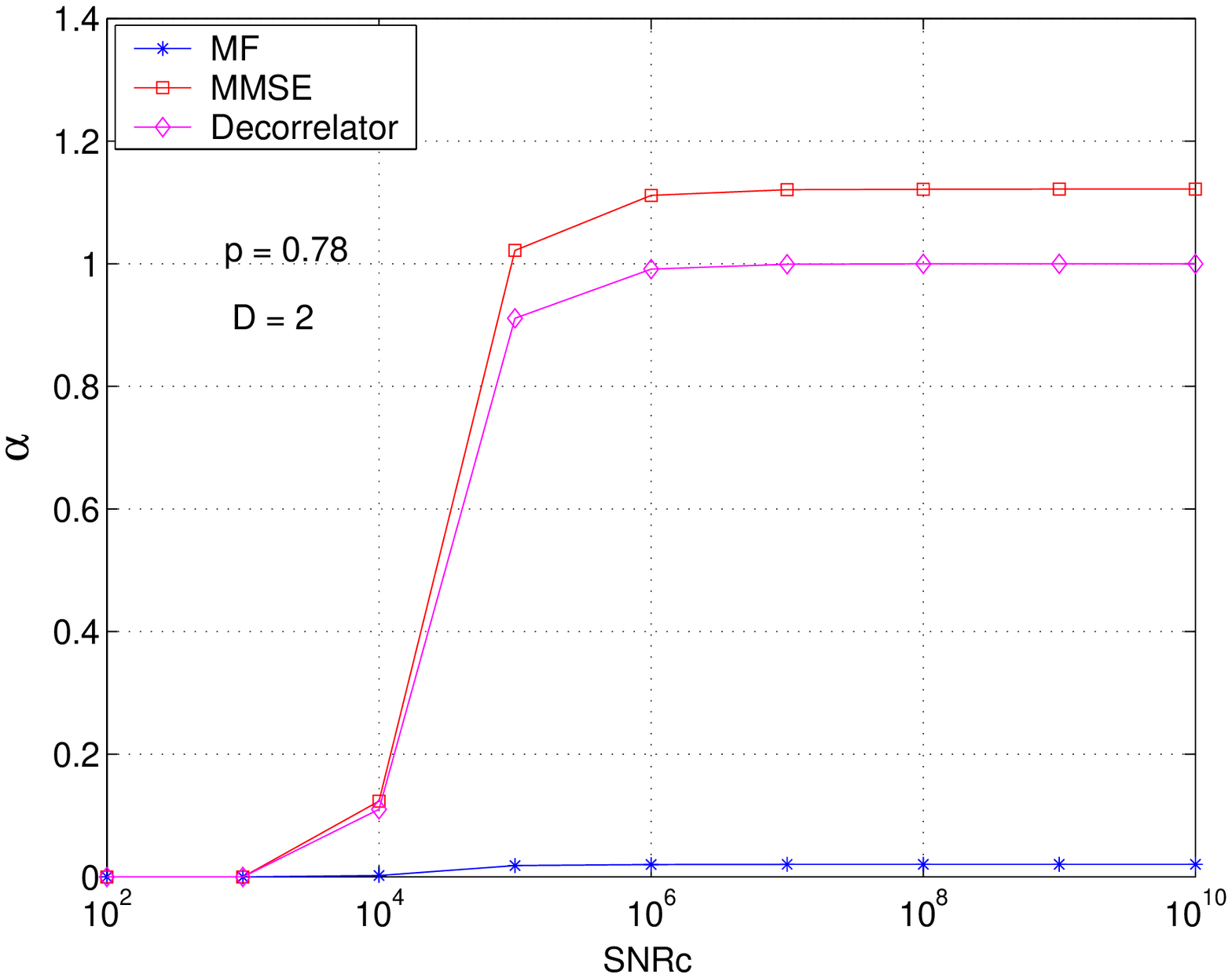}\hspace{.8pc}}
\center{(a)}
\centerline{
\epsfxsize=2.6in\epsffile{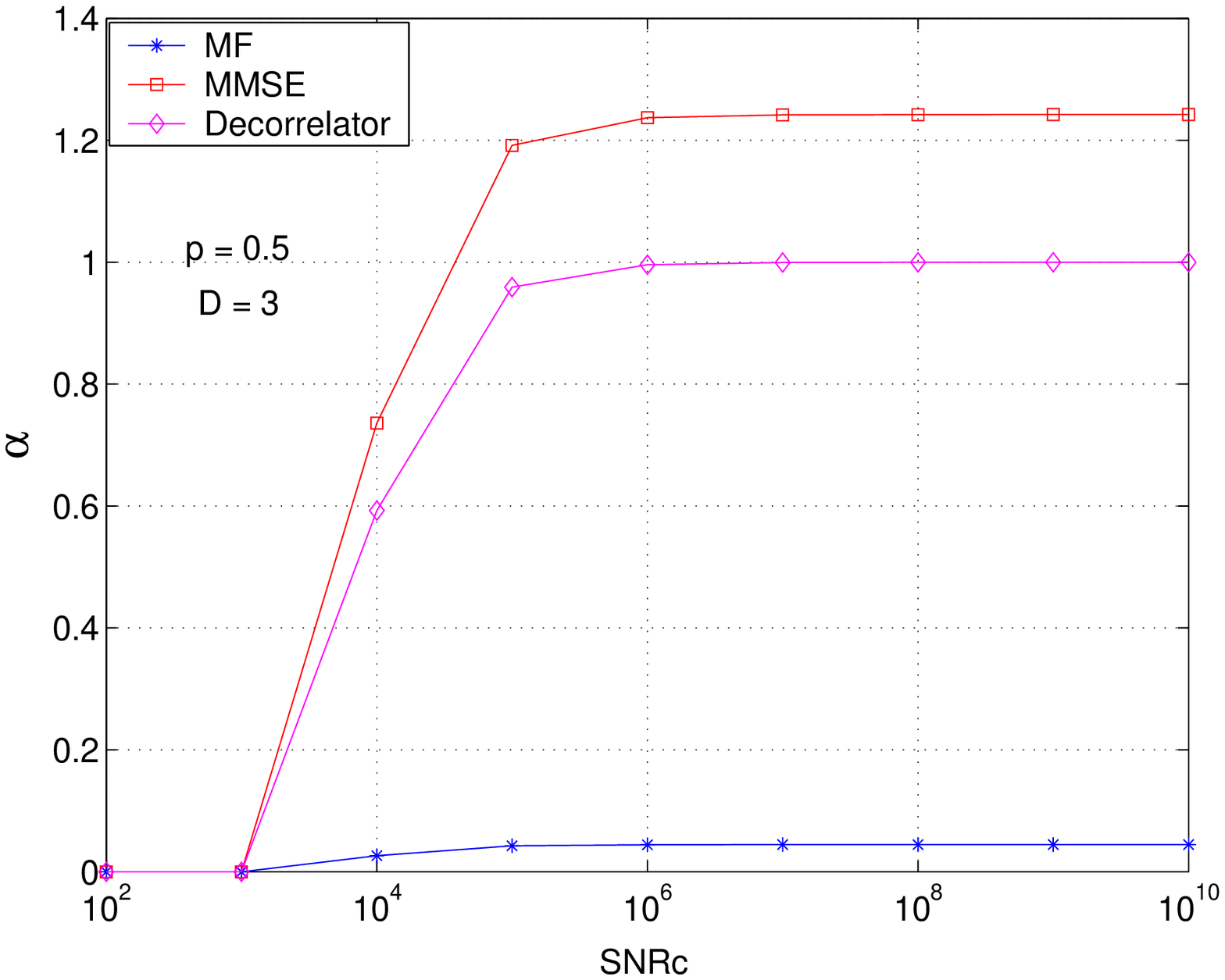}\hspace{.8pc}} 
\center{(b)}
\vspace{-0.3cm}
\caption{Ad hoc network capacity for delay sensitive traffic: (a) D=2; 
(b) D=3.}
\label{fig:capac_sim}
\end{figure}

Figures \ref{fig:capac_sim} (a) and (b) illustrate the network capacity for a 
network diameter constraint of $D=2$ and $D=3$, respectively. Figure 
\ref{fig:capac_sim} shows the number of users per dimension that can be supported 
in an ad hoc network for a given delay constraint, as a function of the maximum 
transmission power requirement,
$SNR_c=\bar{P_t}/{\sigma^2}$. It can be seen that, using multiuser receivers,
almost cellular capacity (obtained for the case with multiuser receivers) can 
be obtained even for very stringent delay ($D=2$, $D=3$) and power 
requirements (transmission power $\bar{P_t}=10^5\sigma^2$).

\subsection{Network Throughput}

As we have seen in the previous subsections, ad hoc network capacity is
greatly enhanced by using a CDMA access method and  separating the users using
multiuser detectors. Tight power and delay constraints can thus be met in 
such networks. We will show now that using multiuser detectors in CDMA ad hoc
networks also improves the overall throughput of the network. To see this, we
compare the network throughput that can be achieved for our analysis by the 
MMSE receiver,  with the scenario described in  \cite{gkumar}, in 
which random access is used. No delay constraints are enforced, and very 
similar network models are used for comparisons: all nodes are randomly 
located and independently and uniformly distributed in a unit area (disc for 
\cite{gkumar}, square in our analysis), each node transmits traffic to a 
randomly chosen destination, all nodes transmit with the same power 
and the transmission rate is $R$. Both synchronous and asynchronous 
transmission cases are considered for the CDMA network and MMSE receivers are
considered.
 
\begin{figure}[ht]
\centerline{
\epsfxsize=2.6 in\epsffile{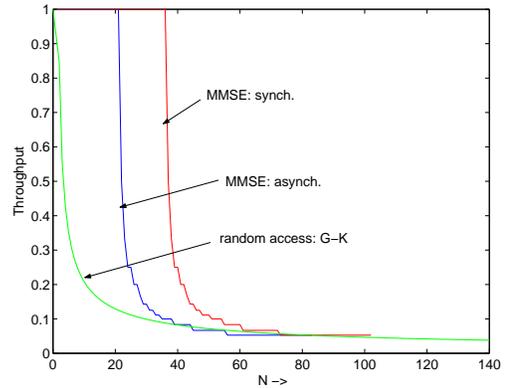}} 
\caption{Network throughput comparison}
\label{fig:gk_comp}
\end{figure}

For the random access scenario, the order of the average throughput 
capacity has been shown in  \cite{gkumar} to be $l(N)=\theta\left( \frac{R}{\sqrt{Nlog(N)}}\right)$. For the CDMA network we approximate the network 
throughput based on (\ref{eq:max_th}): $l(N) \approx \frac{R}{D(N)}$, where 
$R=W/L$. The dependence of the network diameter on the number of nodes can be 
easily determined using (\ref{eq:thr}) and the appropriate formula for $T$ given
the type of receiver used (see Section \ref{section:phy}).
We compare the network throughput for the Gupta-Kumar analysis (G-K) 
\cite{gkumar}, with both a synchronous and an asynchronous CDMA network using 
MMSE receivers.
The same numerical values as before are selected for the example plotted in 
Figure \ref{fig:gk_comp}, which shows the normalized network throughput as a 
function of the number of nodes per unit area. The spreading gain is chosen to 
be $L=32$.

We can see that, although the CDMA ad hoc network capacity also decreases with
the increase of the number of nodes per unit area, its capacity is 
significantly higher than the random access network (G-K).
 Also, the use of MMSE receivers yields unreduced
throughputs for the network for a fairly large network (approx. 
40 nodes per unit area for synchronous transmission). Of course, this advantage 
comes at the price of an increased implementation complexity in acquiring the 
signature sequences for all users and dynamically adjusting the receivers.

\section{Capacity for Finite Networks: Simulations \label{section:sim}}

The capacity results obtained in the previous section are asymptotic in nature,
thus requiring validation through simulations for practical finite networks.
 Since we showed in the previous section 
that the network using matched filters performs poorly compared with a system 
using multiuser detectors, the emphasis is on networks using multiuser
receivers, and the results are only validated for the matched filter case.
All the experiments consider unlimited power transmission for the MMSE case,
and maximum power constraints for the decorrelator, $\bar{P_t}=10^4\sigma^2$ 
(the case of the decorrelator with unlimited transmission power is trivial: 
$\alpha<1$).
For implementation simplicity, all numerical results are obtained for 
synchronous transmission, and using $b=6$, $\lambda = 0.1 \  m$ and $\gamma=5$.
Our experiments consist of selecting a finite (variable) number of 
nodes and randomly generating their locations uniformly across a square area.
Then, the link gains, and consequently the achieved SIRs are computed for all 
pairs of nodes, using Eqs. (\ref{eq:SIR_mf}), (\ref{eq:sirdec}), and 
(\ref{eq:sirmmse}), respectively. 
We note that the simulations do not consider the SIR formulas' accuracy for
finite systems, as this issue has already been studied in \cite{tsezei}, where 
it was shown that the standard deviation for the achieved SIR goes to zero as 
$1/\sqrt{N}$.

If the computed SIR is greater than or equal 
to the target SIR, the link is feasible. The adjacency matrix is then 
 constructed, and based on it, the network diameter is determined. 
The computation of the network diameter uses Dijkstra's algorithm 
\cite{Berts_dn}, as a Matlab function from the Bayes Net Toolbox package 
\cite{BNT}.
The experiment is repeated 100 times and the probabilities associated with 
a range of network diameters are determined. An infinite network diameter
means that the network is disconnected. 

We also determine the probability $p$
of a feasible link  and we compare it with our theoretical results.
Some simulation examples are presented in Tables \ref{tab:resultsdec}, 
\ref{tab:resultsmmse}, and \ref{tab:resultsmf}, for different values of $L$ and $N$,
selected such that we will get a range of values for $\alpha$. It can be seen that 
both the physical layer capacity results, reflected in the achievable link probability 
$p$, as well as the network performance results
(i.e., the achieved network diameter)  are very close to the asymptotic ones, 
especially for larger numbers of nodes in the network cell (the considered 
square area).
\begin{table*} 
\caption{\it Simulation results: Decorrelator}
\vspace{0.1cm}
\begin{center}

\begin {tabular}{|c|c|c|c|c|c|c|} 
\hline 
Receiver & L & N & $p$ (analysis) & $p$ (sim.) & D (asymptotic) & D (sim.) \\ \hline \hline
Decorrelator & 512 & 60 & p= 0.7773 & p=0.7472 &  $D \approx 2$  &
                \epsfxsize=1in
                \epsfbox{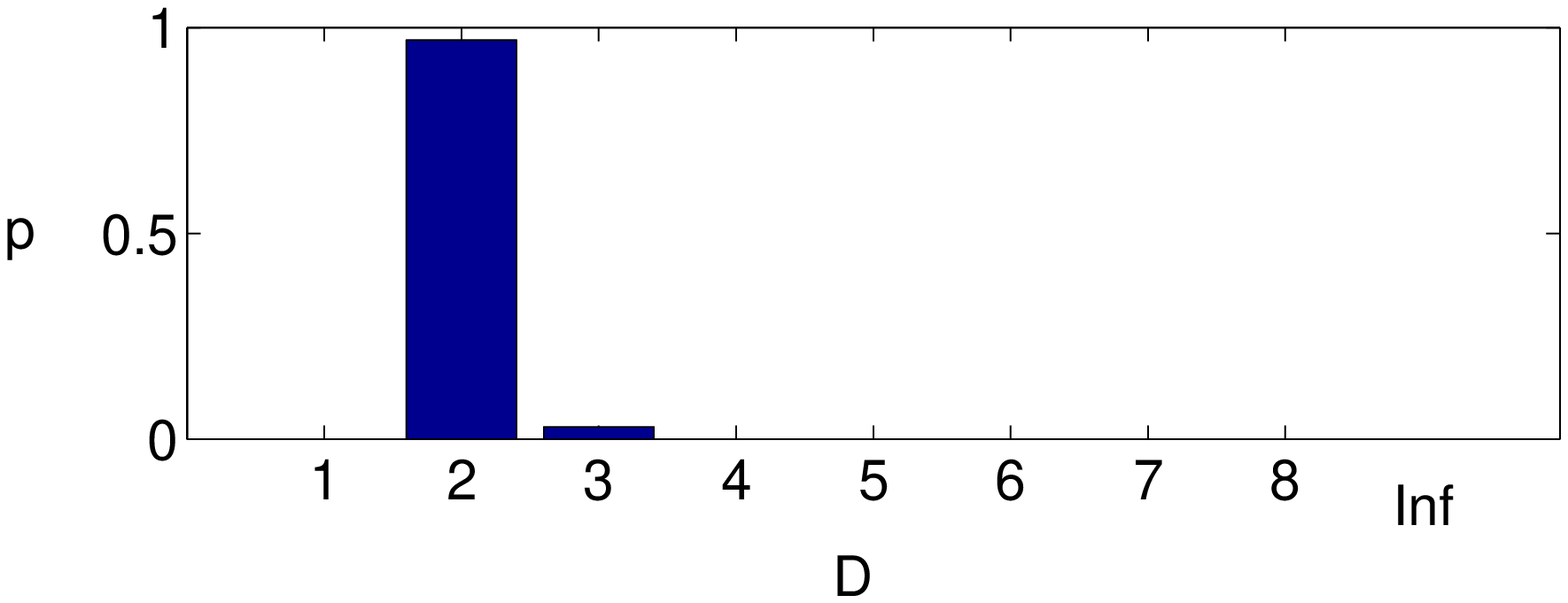} \\ \hline
Decorrelator & 1024 & 120 & p=0.7773  & p= 0.7510&  $D \approx 2$  &
                \epsfxsize=1in
                \epsfbox{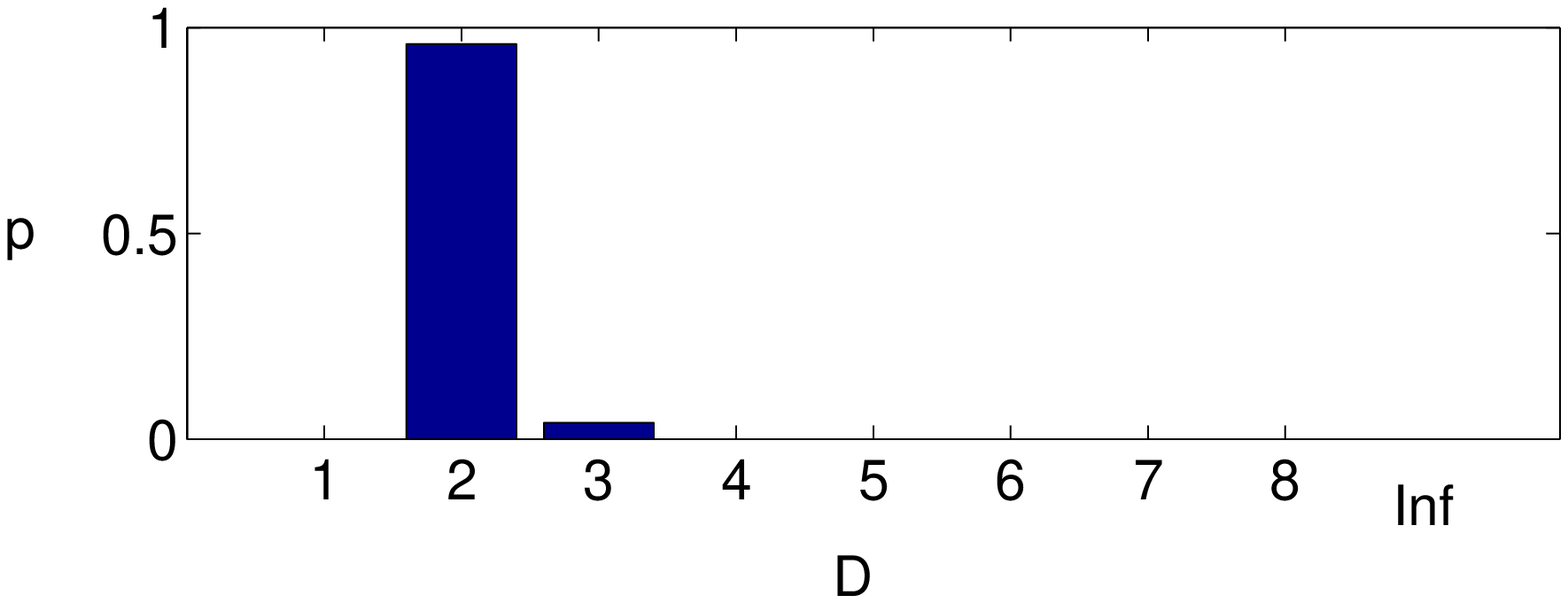} \\ \hline
Decorrelator & 64 & 28 & p=0.6160 & p=0.5670 & $D \approx 3$ &
                \epsfxsize=1in
                \epsfbox{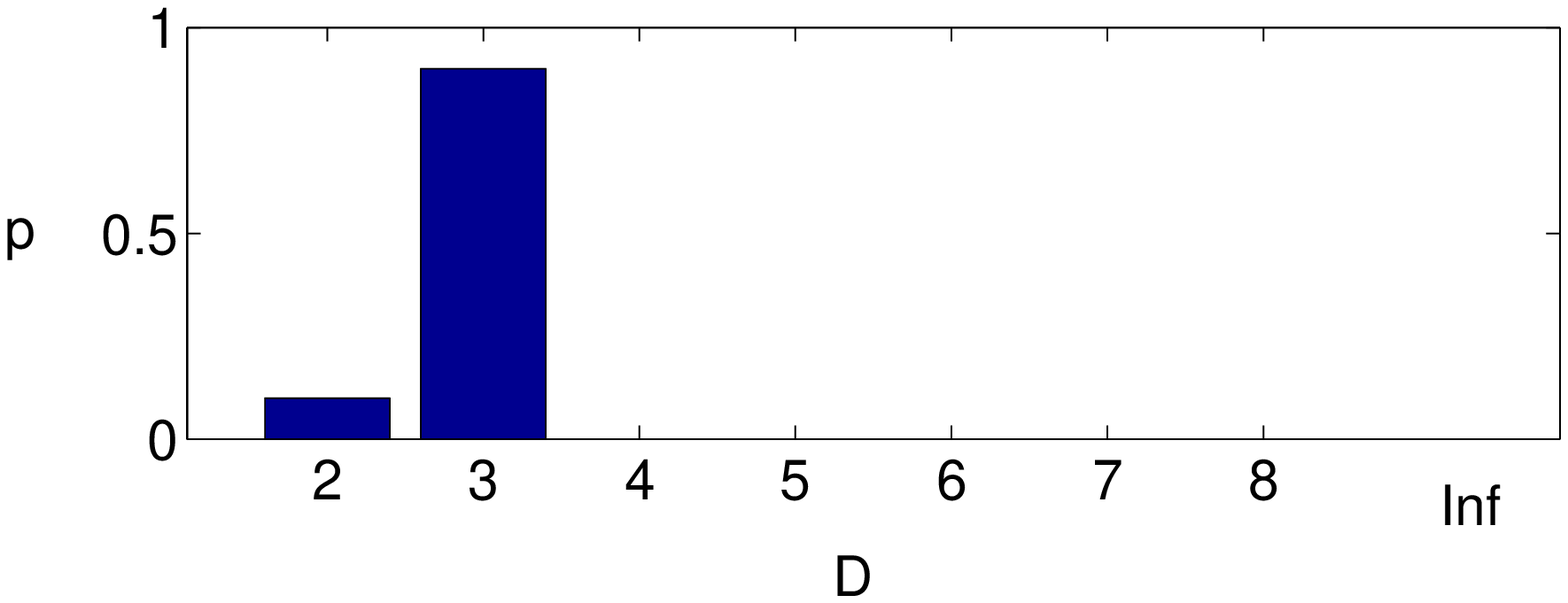} \\  \hline
Decorrelator & 128 & 92 & p=0.3803 & p=0.3392 & $D \approx 4$ &
                \epsfxsize=1in
                \epsfbox{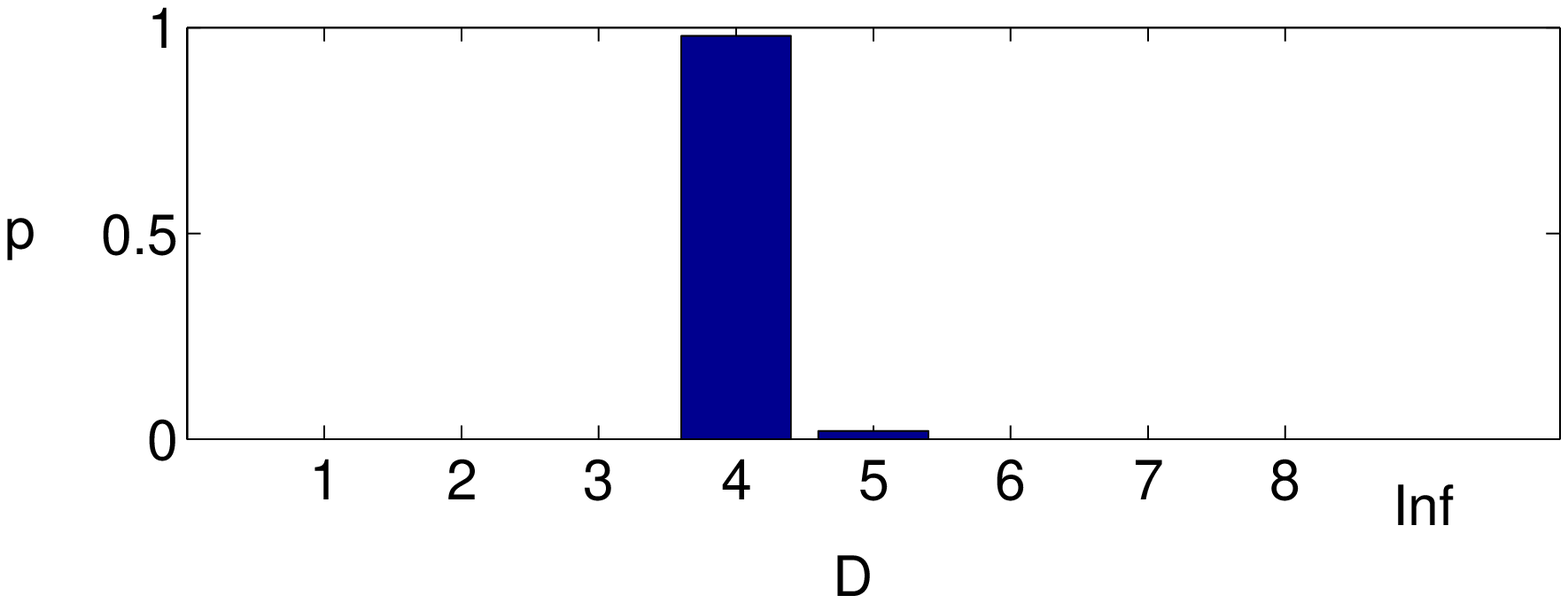} \\  \hline
Decorrelator & 128 & 96 & p=0.3464 & p=0.3074 & $D \approx 4$ &
                \epsfxsize=1in
                \epsfbox{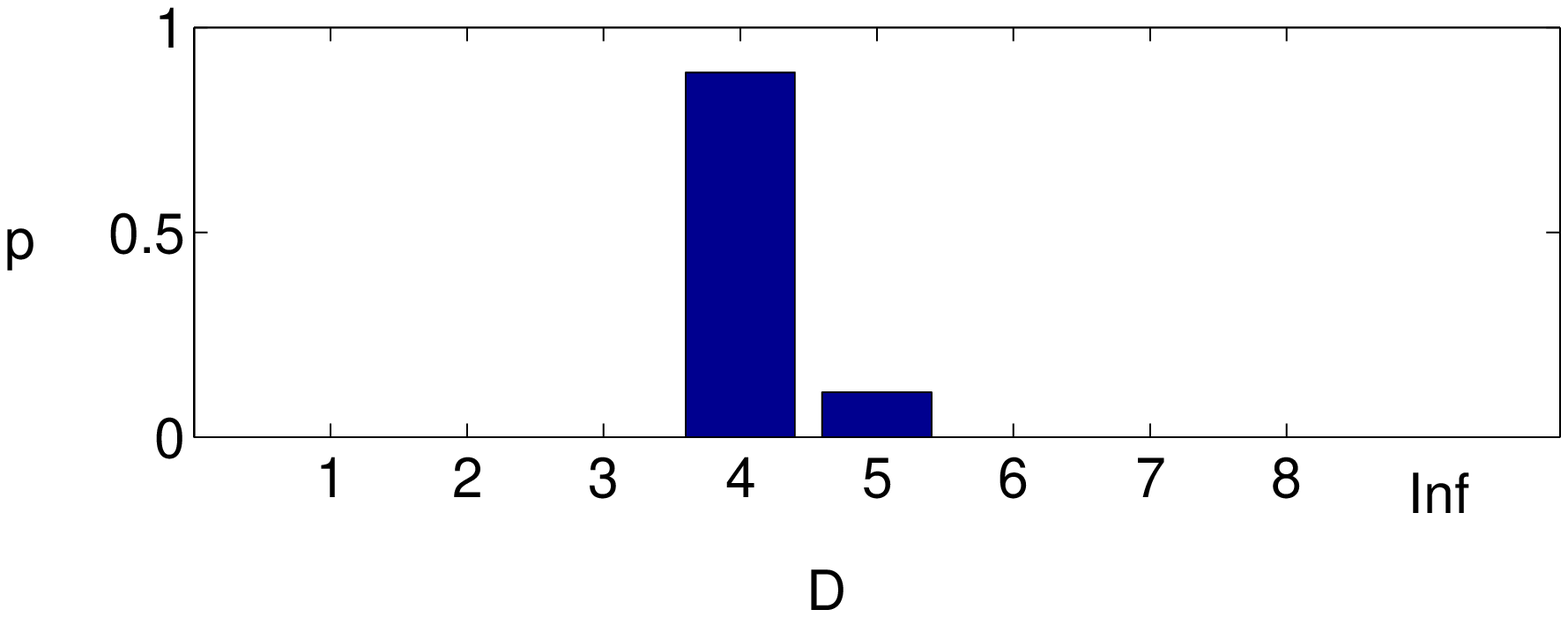} \\  \hline
Decorrelator & 128 & 100 & p=0.3107 & p=0.2764 & $D \approx 4$ &
                \epsfxsize=1in
                \epsfbox{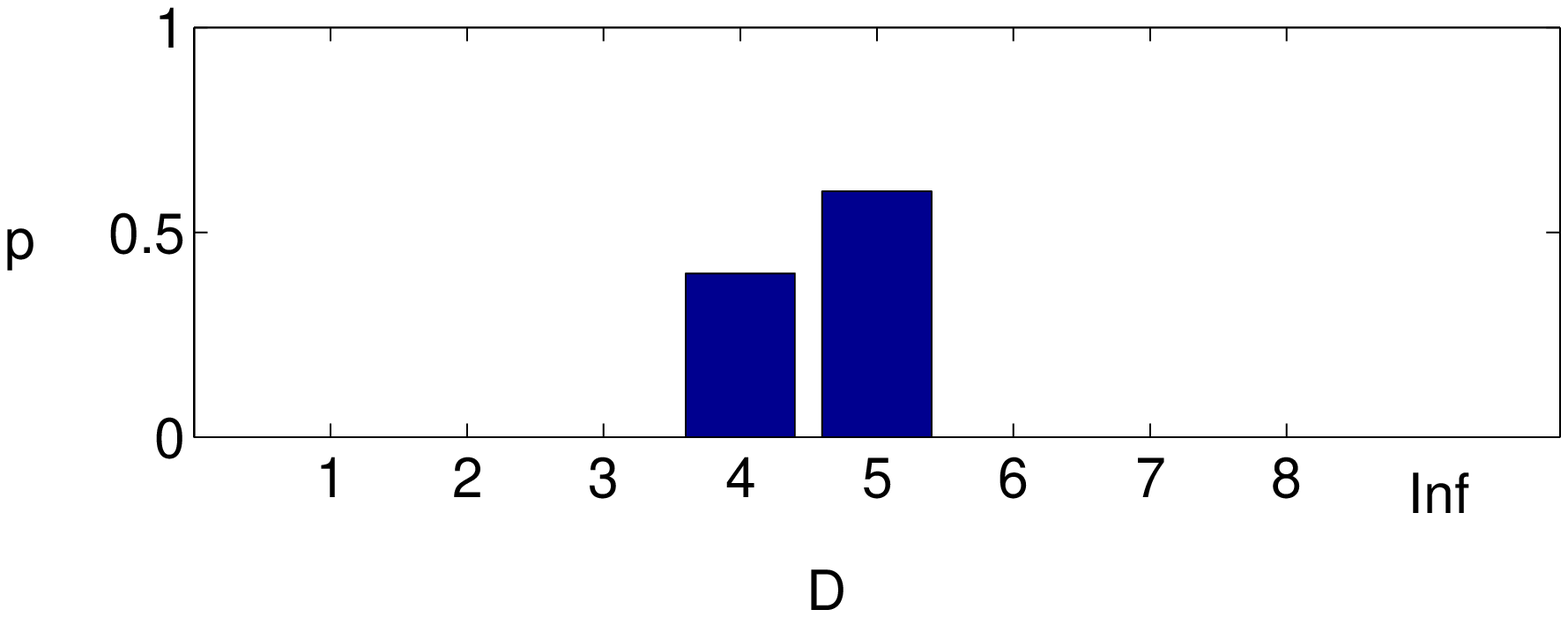} \\  \hline
Decorrelator & 64 & 57 & p=0.1698 & p=0.1515 & $D \approx 7$ &
                \epsfxsize=1in
                \epsfbox{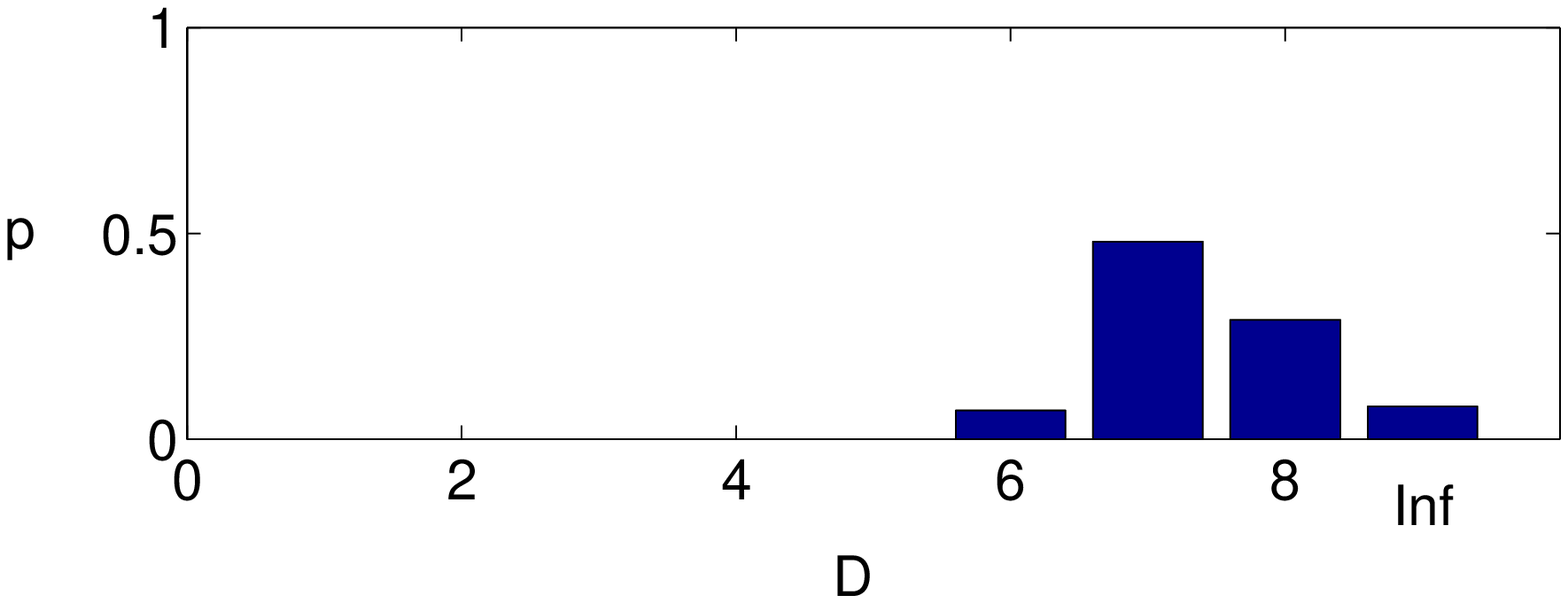} \\  \hline
\hline
\end{tabular}
\end{center}
\label{tab:resultsdec}
\vspace*{-0.4cm}
\end{table*} 
\begin{table*} 
\caption{\it Simulation results: MMSE}
\vspace{0.1cm}
\begin{center}
\begin {tabular}{|c|c|c|c|c|c|c|} 
\hline 
Receiver & L & N & $p$ (analysis) & $p$ (sim.) & D (asymptotic) & D (sim.) \\ \hline \hline

MMSE & 32  & 38 & p=0.6056 & p=0.7491 &  $D \approx 2$  &
                \epsfxsize=2in
                \epsfbox{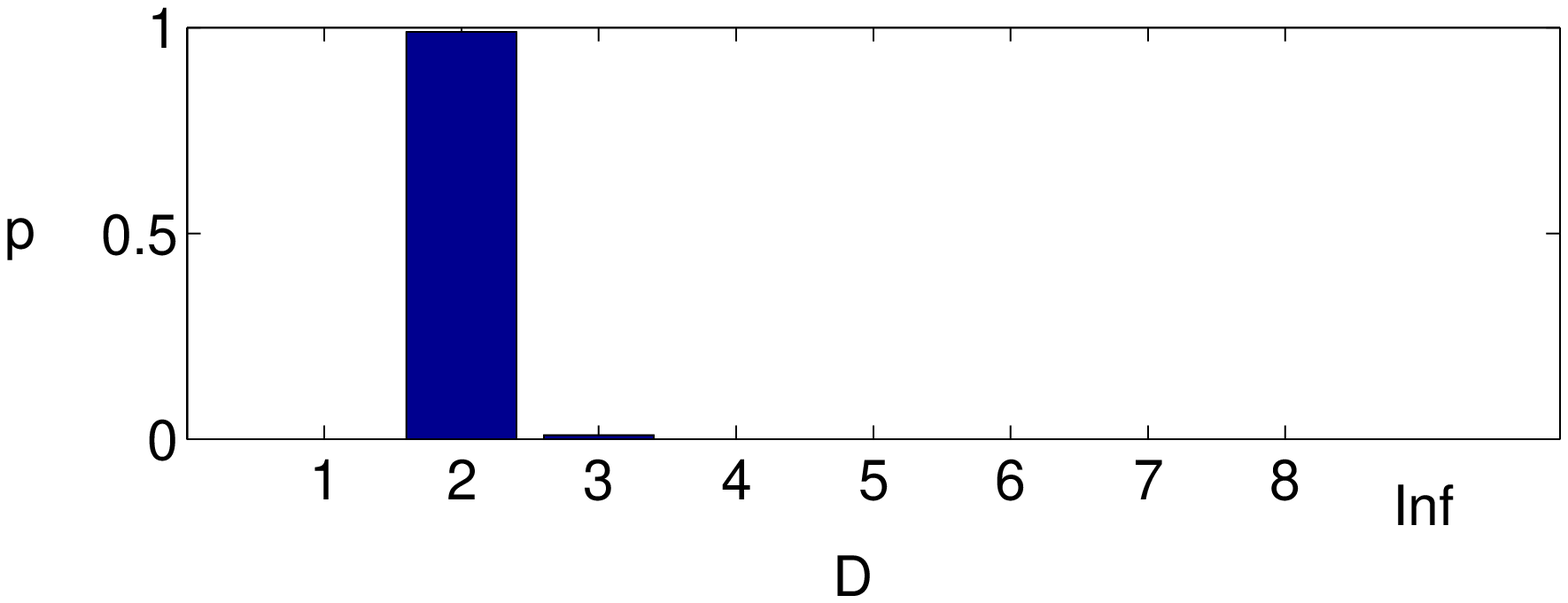} \\ \hline
MMSE & 32  & 39 & p=0.5415 & p=0.4886 & $D \approx 3$ &
                \epsfxsize=2in
                \epsfbox{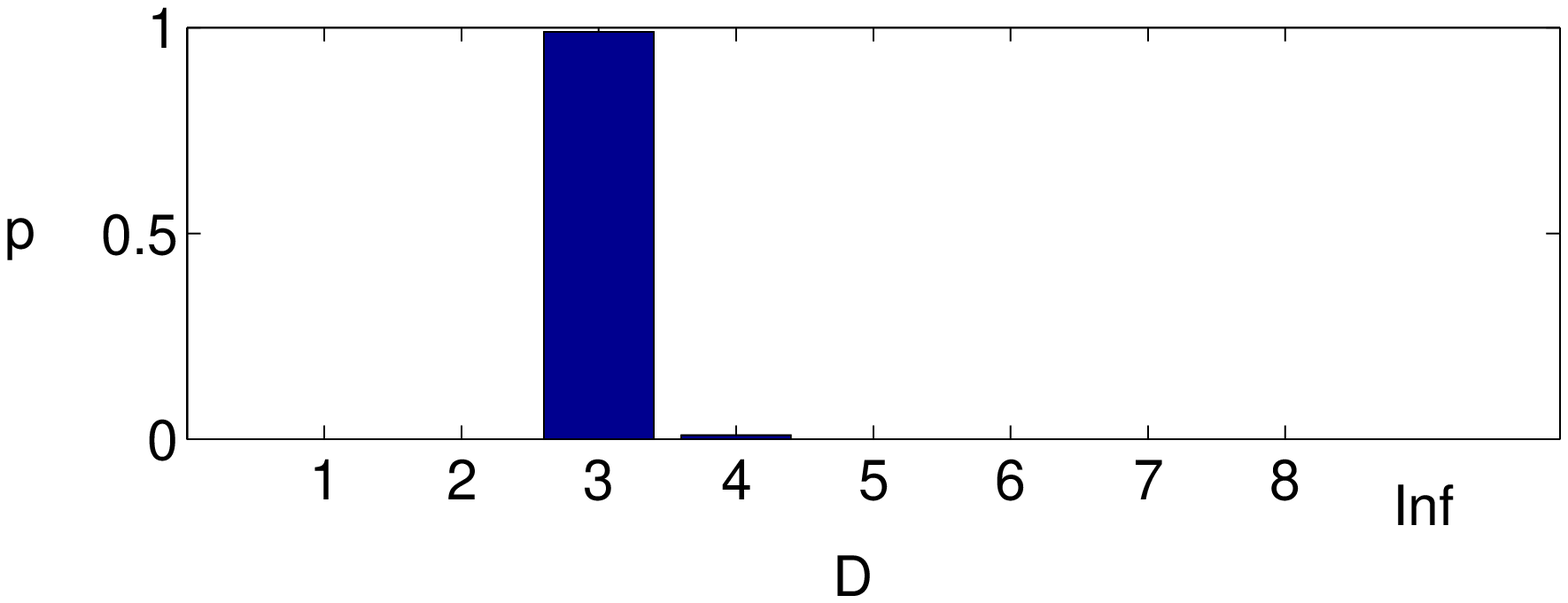} \\  \hline
MMSE & 32  & 42 & p=0.4024 & p=0.433 & $D \approx 3/4$ &
                \epsfxsize=2in
                \epsfbox{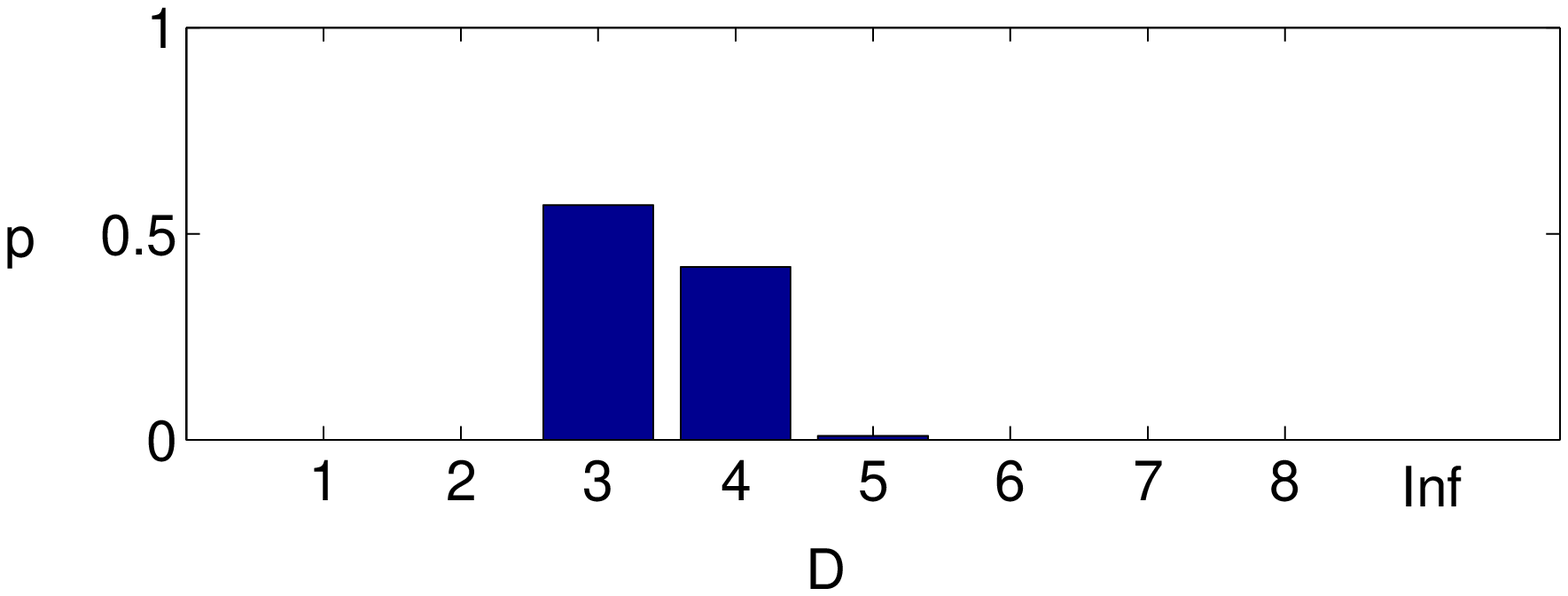} \\  \hline
MMSE & 32  & 45 & p=0.3137 & p=0.3260 & $D\approx 4$ &
                \epsfxsize=2in
                \epsfbox{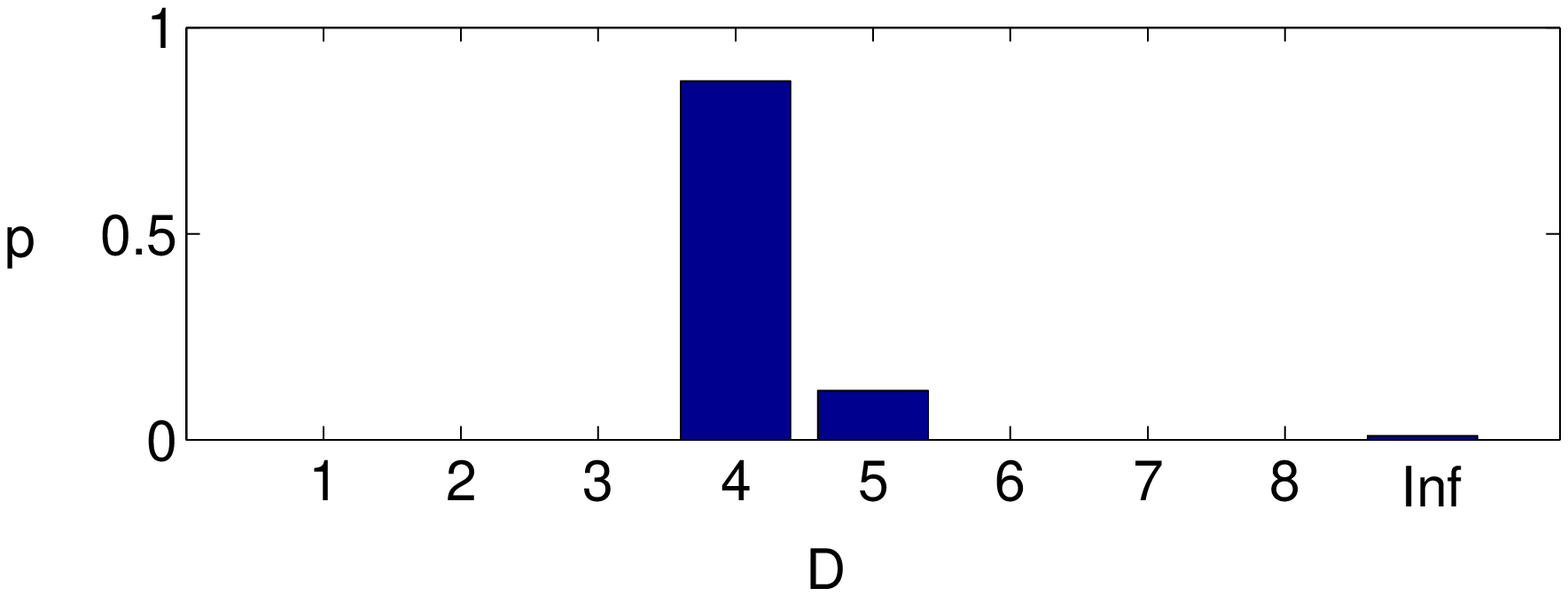} \\  \hline
MMSE & 32  & 46 & p=0.2913 & p=0.2983 & $D \approx 4$ &
                \epsfxsize=2in
                \epsfbox{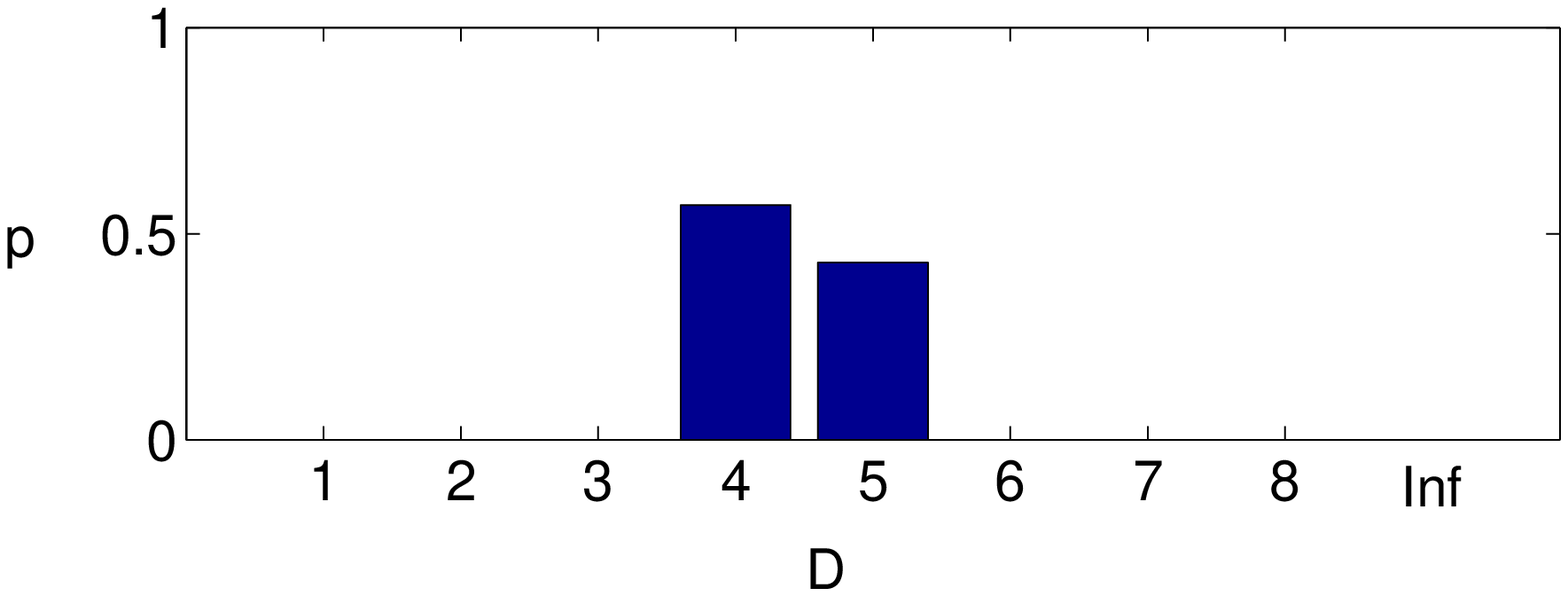} \\  \hline
MMSE & 32  & 48 & p=0.2537 & p=0.2590 & $D \approx 5$ &
                \epsfxsize=2in
                \epsfbox{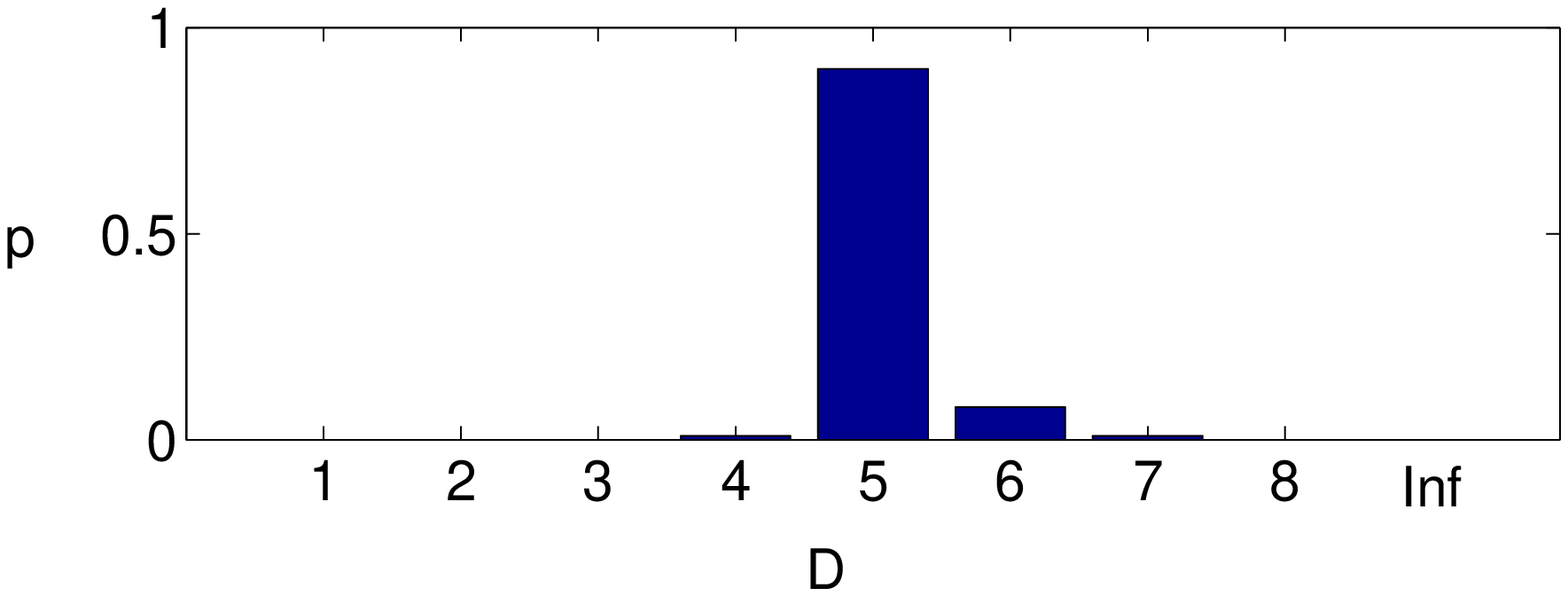} \\  \hline
MMSE & 32  & 57 & p=0.1546 & p=0.1584 & $D \approx 7$ &
                \epsfxsize=2in
                \epsfbox{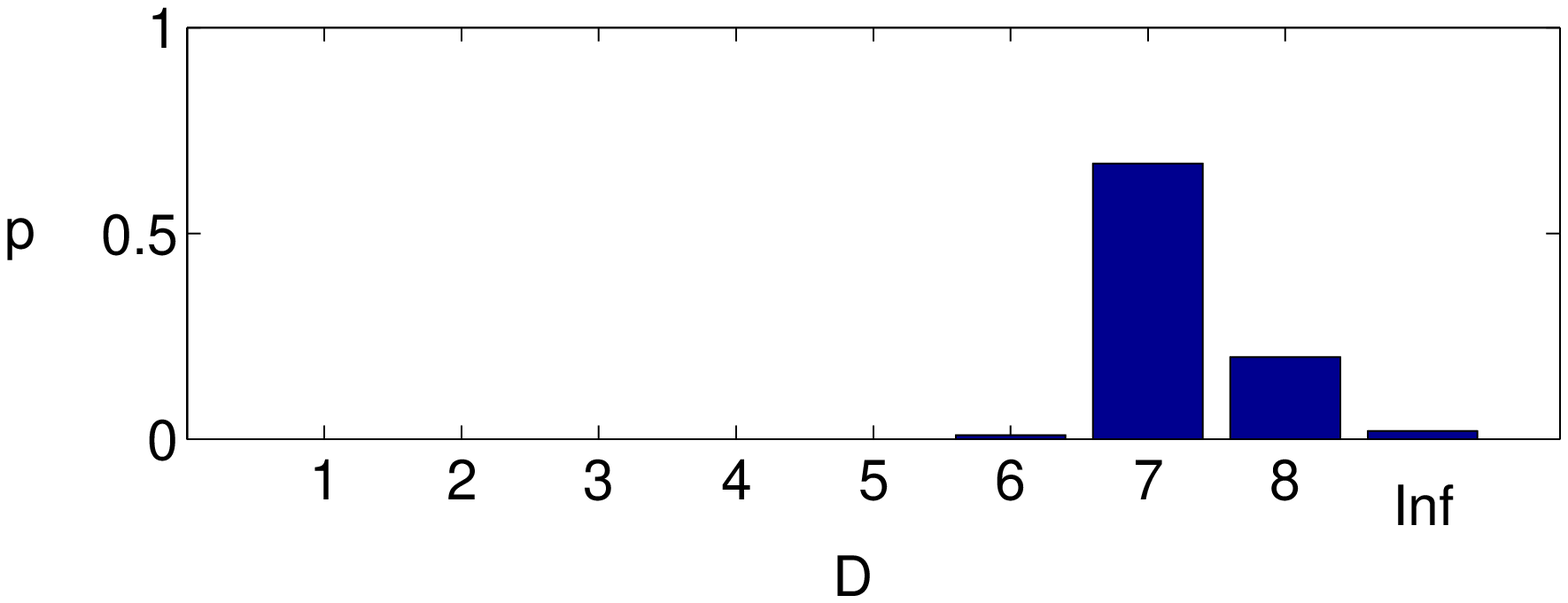} \\  \hline
MMSE & 64  & 78 & p=0.5415 & p=0.5490 & $D \approx 3$ &
                \epsfxsize=2in
                \epsfbox{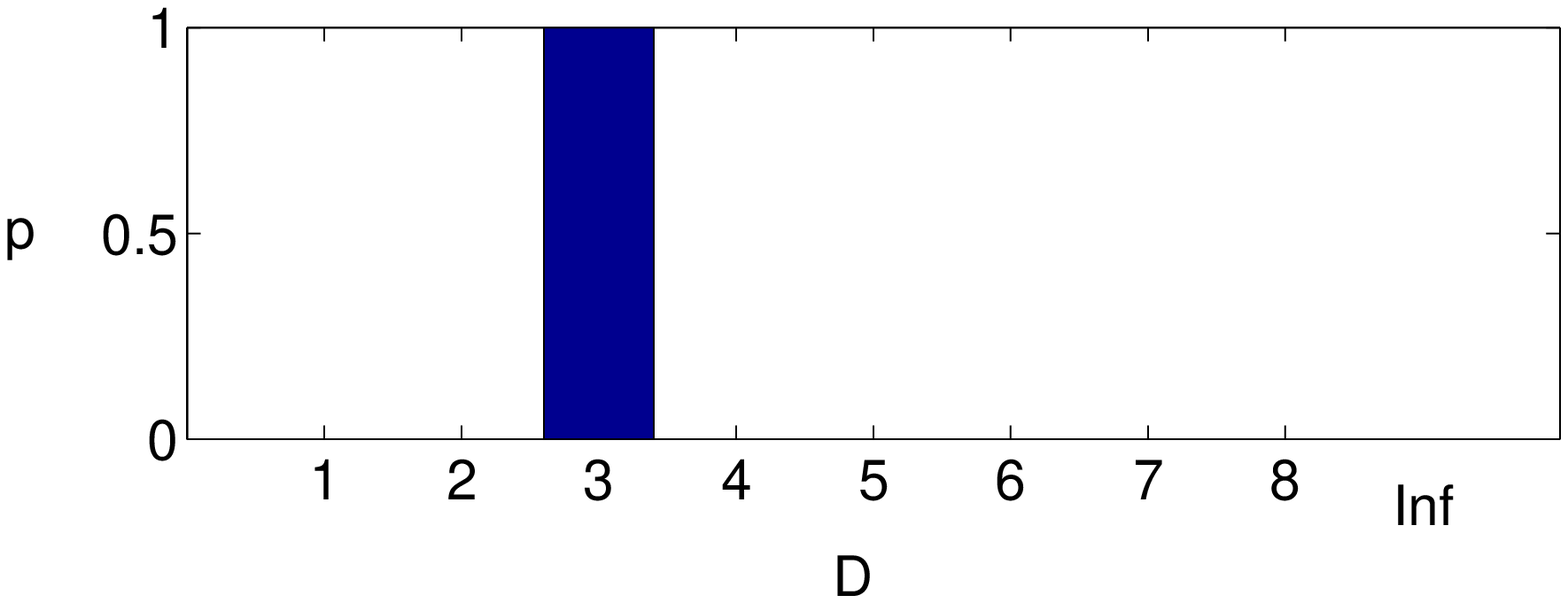} \\  \hline
MMSE & 64  & 74 & p=0.6814 & p=0.7482 &  $D \approx 2$  &
                \epsfxsize=2in
                \epsfbox{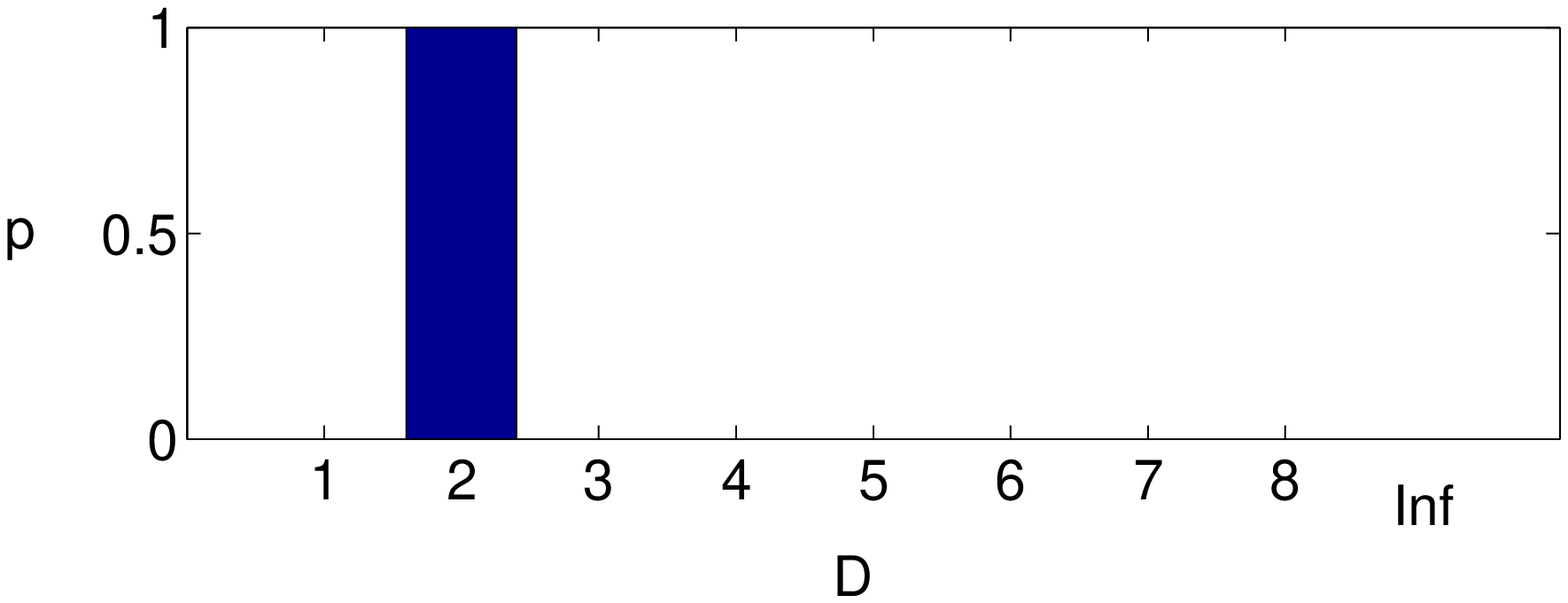} \\ \hline

\hline
\end{tabular}
\end{center}
\label{tab:resultsmmse}
\vspace*{-0.4cm}
\end{table*} 

\begin{table*} 
\caption{\it Simulation results: MF}
\vspace{0.1cm}
\begin{center}
\begin{tabular}{|c|c|c|c|c|c|c|} 
\hline 
Receiver & L & N & $p$ (analysis) & $p$ (sim.) & D (asymptotic) & D (sim.) \\ \hline \hline
MF & 1024 & 44 & p=0.5117 & p=0.6107 & $D \approx 3$ &
                \epsfxsize=2in
                \epsfbox{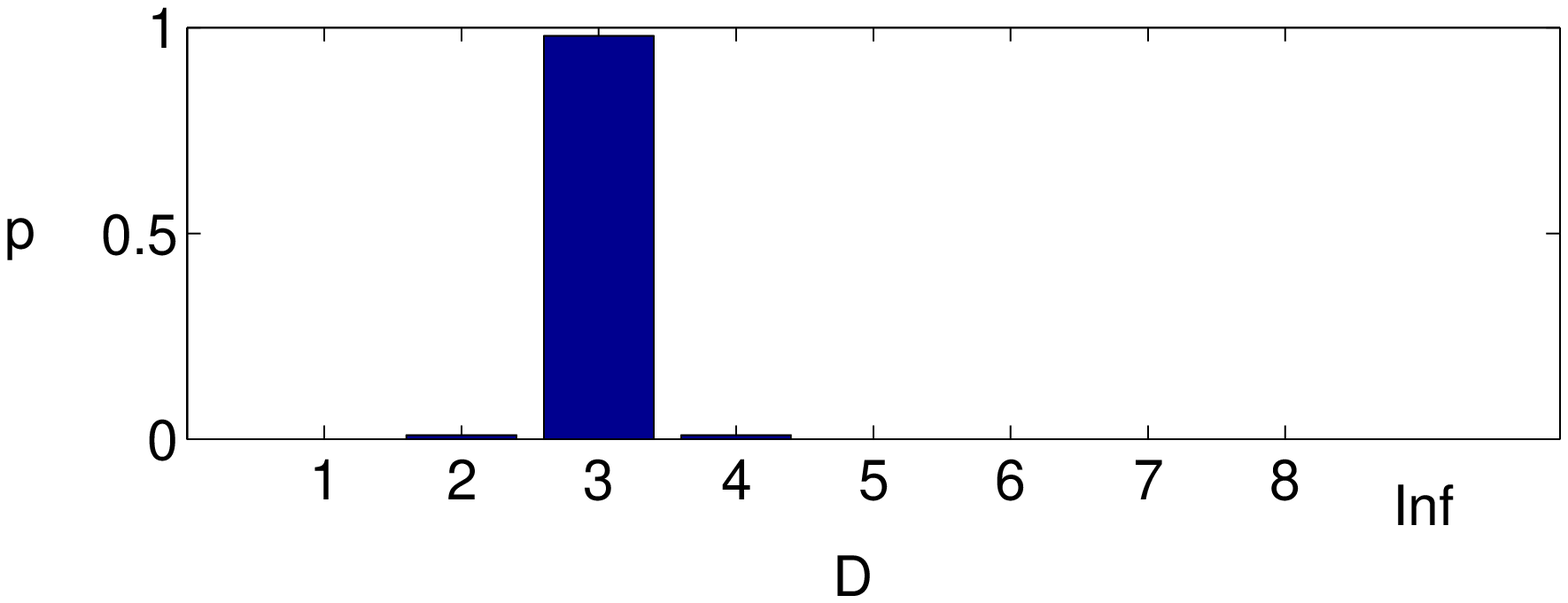} \\ \hline
MF & 256 & 31 & p=0.2246 & p=0.3093 & $D \approx 5$ &
                \epsfxsize=2in
                \epsfbox{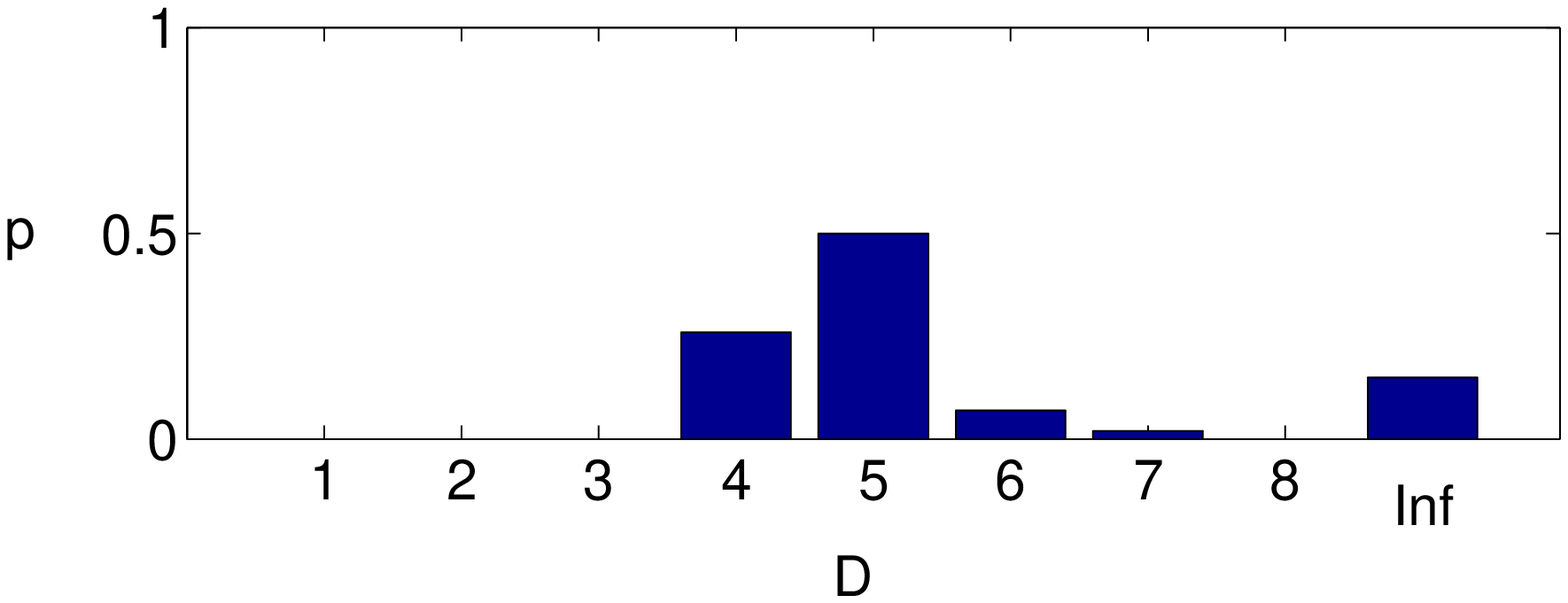} \\ \hline
MF & 512 & 144 & p=0.1037 & p=0.1127 & $D \approx 8$ &
                \epsfxsize=2in
                \epsfbox{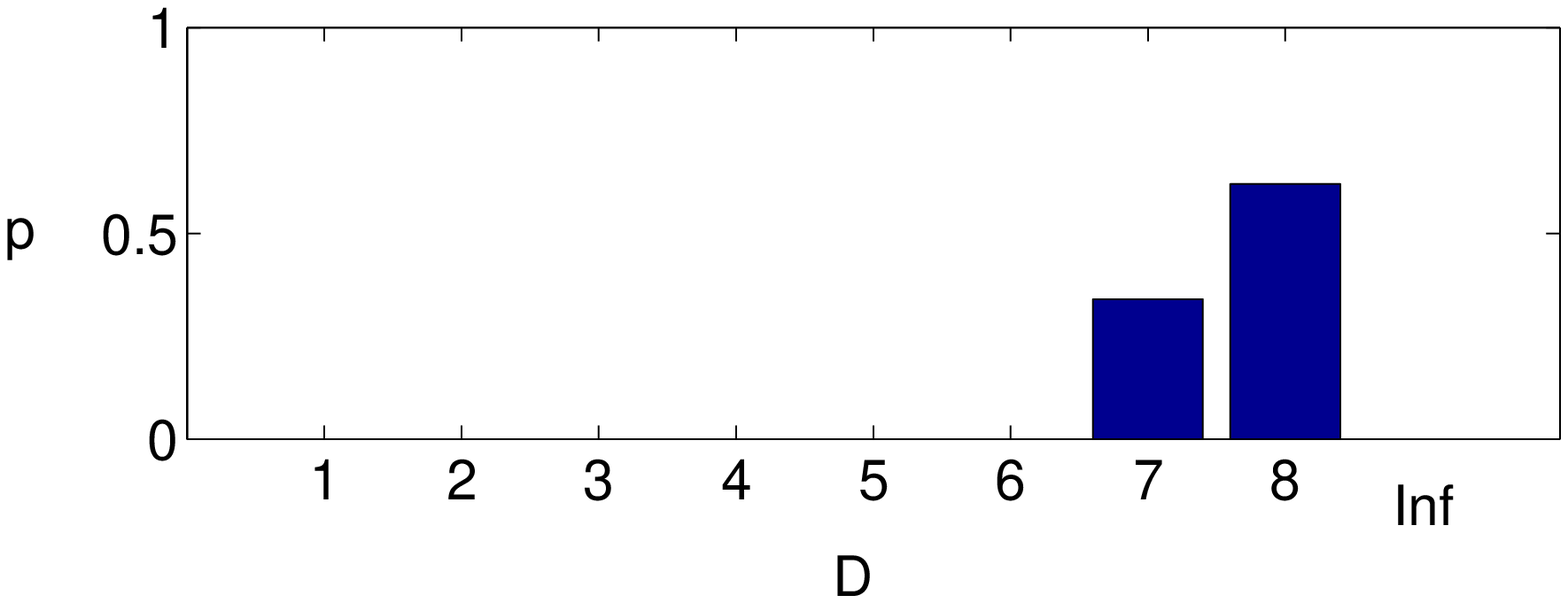} \\ \hline 
\hline
\end{tabular}
\end{center}
\label{tab:resultsmf}
\vspace*{-0.4cm}
\end{table*}

\section{Conclusions}

In this paper we have analyzed the asymptotic capacity for delay
sensitive traffic in ad hoc networks. While previous results have 
focused on enhancing the network capacity at the expense of increased 
transmission delay, our approach is to exploit advanced signal processing 
techniques, such as multiuser detection, to enhance capacity when tight delay 
constraints are enforced. We have analyzed three different network scenarios 
for a DS-CDMA air interface in which the users have matched filters,
decorrelating or MMSE receivers. We combined physical layer requirements
(signal to interference ratio) with network layer QoS constraints
(transmission delay). The maximum network transmission delay has been expressed
in terms of the maximum number of hops for any arbitrarily selected 
source-destination pair of nodes. We then have characterized the network delay 
using geometric arguments for the asymptotic case.
Since all derivations in this paper are asymptotic in nature, simulation 
results have been presented for performance validation with finite systems. 
Both analysis and simulations have shown significant network capacity gains
for ad hoc networks employing multiuser detectors, compared with those using
 matched filters, as well as very good performance even under tight delay and 
power constraints.

\nocite{tong_merg}
\nocite{sank}
\nocite{rodop}
\nocite{silv_sousa}

\bibliography{ref_adhoc}
\bibliographystyle{abbrv}

\centerline{\large \bf Appendix}

\vspace{.3cm}
{\bf Normalized conditional average interference derivation for MMSE networks:}
\vspace{.5cm}

Given the fact that the link gain $h$ takes
values in the interval $\left[\lambda^2/\delta_m^2,  \ \lambda^2/\delta_M^2 \right]$ with high probability, the normalized conditional average interference derivation for MMSE networks can be approximated as: 

\[ E[H|h_i]\approx C \int_{1/\delta_M^2}^{1/\delta_m^2}\frac{h_i}{h(h_i+h\gamma)}\exp\left(-\frac{C}{h} \right)dh.\]

\noindent Denoting $x=C/h$, we have
\[ E[H|h_i] \approx C \int_{\delta_m^2C}^{\delta_M^2C}\frac{h_i}{h_ix+C\gamma}\exp(-x)dx.\]

\noindent Again, denoting $y=h_ix+C\gamma$, we further have
\[ E[H|h_i] \approx C \exp\left(\frac{C \gamma}{h_i}\right)\int_{h_i\delta_m^2C+C \gamma}^{h_i\delta_M^2C+C \gamma}\frac{1}{y}\exp\left(-\frac{y}{h_i} \right)dy.\]

\noindent On making a further change of variable $z=y/h_i$, we arrive at
\[ E[H|h_i] \approx C \exp\left(\frac{C \gamma}{h_i}\right)\int_{\delta_m^2C+ (C \gamma)/h_i}^{\delta_M^2C+(C \gamma)/h_i}\frac{1}{z}\exp\left(-z \right)dz,\]

\noindent which yields 

\[E[H|h_i] \approx C\exp\left(\frac{C \gamma}{h_i}\right) \left[E_1\left(\delta_m^2C+\frac{C \gamma}{h_i}\right)-E_1\left(\delta_M^2C+\frac{C \gamma}{h_i}\right)  \right], \]

\noindent where $E_1(x)=\int_x^{\infty}\frac{1}{t}\exp(-t)dt$  is the exponential integral.

\end{document}